\documentclass[a4paper,11pt]{article}
\pdfoutput=1
\usepackage{jcappub}
\usepackage{slashed}

\usepackage{mathtools}
\usepackage[T1]{fontenc}
\usepackage[normalem]{ulem}

\graphicspath{{plots/}}

\newcommand{\gs}{g_\star}
\newcommand{\gss}{g_{\star s}}
\newcommand{\Trh}{T_{\rm rh}}
\newcommand{\Tmax}{T_{\rm max}}
\newcommand{\np}{n_\phi}
\newcommand{\ndm}{n_\chi}
\newcommand{\nrh}{n_N}
\newcommand{\rp}{\rho_\phi}
\newcommand{\rR}{\rho_R}
\newcommand{\Gp}{\Gamma_\phi}
\newcommand{\GN}{\tilde\Gamma_N}

\newcommand{\rrh}{\rho_N}
\newcommand{\Br}{\text{Br}}
\newcommand{\aph}{a_\phi}
\newcommand{\anr}{a_{\rm nr}}
\newcommand{\arh}{a_{\rm rh}}

\title{Seesaw Cosmology}
\author[a]{Nicolás Bernal,}
\author[b]{Chee Sheng Fong}
\author[c]{and Óscar Zapata}

\affiliation[a]{New York University Abu Dhabi\\
PO Box 129188, Saadiyat Island, Abu Dhabi, United Arab Emirates}
\affiliation[b]{Centro de Ciências Naturais e Humanas,
Universidade Federal do ABC\\
09.210-170, Santo André, SP, Brazil}
\affiliation[c]{Instituto de Física, Universidad de Antioquia\\
Calle 70 \# 52-21, Apartado Aéreo 1226, Medellín, Colombia}

\emailAdd{nicolas.bernal@nyu.edu}
\emailAdd{sheng.fong@ufabc.edu.br}
\emailAdd{oalberto.zapata@udea.edu.co}

\abstract{We study perturbative reheating in which the inflaton transfers its energy to the Standard Model through right-handed neutrinos (RHNs) responsible for light-neutrino masses via the type-I seesaw mechanism. We refer to the resulting nonstandard thermal history as \emph{seesaw cosmology}. When produced relativistically and sufficiently long lived, the RHNs generate a characteristic sequence of inflaton, relativistic-RHN, nonrelativistic-RHN, and Standard Model radiation domination. We solve the Boltzmann system while retaining the production-time dependence of the nonthermal RHN distribution and its relativistic-to-nonrelativistic transition. The Standard Model temperature rapidly approaches a plateau during inflaton domination and subsequently scales as $a^{-1/4}$ and $a^{-3/8}$ during relativistic- and nonrelativistic-RHN domination, respectively. We investigate the implications of seesaw cosmology for dark-matter production. Direct production through inflaton decays can be enhanced relative to conventional reheating by a factor of order $m_\phi/(2m_N)$, while ultraviolet freeze-in exhibits the critical temperature powers $p = 12$ and $20$, leading to potentially large contributions before the final radiation-dominated era. Seesaw cosmology therefore connects neutrino-mass generation, the pre-BBN thermal history and phenomena such as dark-matter production and baryogenesis.}
\begin{document}
\begin{flushright}
\end{flushright}
\maketitle

%%%%%%%%%%%%%%%%%%%%%%%%%%%%%%%%%%%%%
\section{Introduction}
%%%%%%%%%%%%%%%%%%%%%%%%%%%%%%%%%%%%%
The standard thermal history of the Universe is observationally tested from Big Bang nucleosynthesis onward, whereas its evolution at earlier times remains largely unconstrained~\cite{Allahverdi:2020bys, Batell:2024dsi}. In particular, after inflation, the energy density stored in the field dominating the Universe must be transferred to particles that eventually form the Standard Model (SM) thermal bath. This reheating process establishes the initial conditions for the radiation-dominated era and can strongly affect the production of cosmological relics, baryogenesis, and primordial gravitational-wave backgrounds~\cite{Dolgov:1989us, Traschen:1990sw, Kofman:1994rk, Kofman:1997yn, Amin:2014eta, Barman:2025lvk}. Its dynamics depends not only on the decay rate of the dominant field, but also on the masses, lifetimes, and momentum distributions of the intermediate particles through which energy is transferred.

Right-handed neutrinos (RHNs) provide a particularly economical portal between the post-inflationary Universe and the SM. Inflaton decays into RHNs have been studied in connection with nonthermal leptogenesis and neutrino reheating~\cite{Lazarides:1990huy, Asaka:1999jb, Zhang:2023oyo, You:2024hit, Han:2024qbw, Zhang:2025jfh, Datta:2025wfh}. Relativistic RHNs produced during reheating may retain information about their production process through their redshifted momentum distributions and time-dilated decays~\cite{Jaeckel:2020oet}. In the type-I seesaw mechanism~\cite{Minkowski:1977sc, Yanagida:1979as, Glashow:1979nm, Gell-Mann:1979vob, Mohapatra:1979ia}, their Majorana masses and Yukawa interactions account for the smallness of the observed active-neutrino masses~\cite{Esteban:2024eli, deSalas:2020pgw}. The same particles can also mediate reheating through the sequence
\begin{equation}
    \phi \to N\, N\,,
    \qquad
    N \to \ell\, H\,,
\end{equation}
where $\phi$ denotes the scalar field dominating the energy density after inflation, $N$ the RHN, $\ell$ the SM lepton doublet, and $H$ the Higgs doublet. In this scenario, the parameters governing reheating are connected to neutrino physics: the RHN decay widths depend on the seesaw scale and on the active-neutrino mass spectrum, with lower bounds determined by the measured neutrino mass-squared differences. For RHN widths close to these lower bounds, the post-inflationary history is therefore considerably more predictive than in a generic reheating scenario.

Cosmologies in which heavy RHNs generate the SM radiation bath have been considered in different contexts, including entropy production, RHN-assisted reheating, and dark-matter production~\cite{Buchmuller:2011mw, Haque:2023zhb, Cosme:2024ndc, Haque:2024zdq, Coy:2024itg, Borboruah:2025hai, Datta:2025wfh, Mambrini:2026tla}. Inflaton decays into RHNs are also central to nonthermal leptogenesis~\cite{Lazarides:1990huy, Asaka:1999yd, Hahn-Woernle:2008tsk}, while their possible gravitational-wave signatures have recently attracted considerable attention~\cite{Ringwald:2020ist, Ghoshal:2022kqp, Barman:2023ymn, Barman:2023rpg, Bernal:2023wus, Barman:2024htg, Bernal:2025lxp, Cline:2026jra}. In many existing treatments, however, the RHNs are assumed to be nonrelativistic throughout the relevant evolution, or the inflaton is allowed to decay directly into both RHNs and SM radiation. This assumption need not hold when $m_\phi \gg 2\, m_N$. In that regime, RHNs are produced with momenta of order $m_\phi/2$, and can remain relativistic for a long period before their momenta redshift below their mass. If they are sufficiently long lived, they may dominate the energy density first as a radiation-like component and subsequently as a matter-like component before decaying into the SM bath.

In this work, we study this two-step reheating process while retaining the full relativistic-to-nonrelativistic evolution of the nonthermal RHN population. Because the inflatons decay continuously, RHNs produced at different times have different momenta, time-dilated decay rates, and equations of state at a given epoch. The system therefore cannot, in general, be described by assigning a single average momentum or decay rate to the entire RHN population. We instead resolve the distribution according to the RHN production time and reconstruct its number density, energy density, and pressure by integrating over all production cohorts.

The resulting cosmological history, which we refer to as \emph{seesaw cosmology}, can contain four successive regimes. The Universe is initially dominated by the coherently oscillating inflaton, with an effective equation of state $\omega \simeq 0$. It subsequently becomes dominated by relativistic RHNs, for which $\omega \simeq 1/3$. The RHNs then redshift to the nonrelativistic regime, restoring a matter-dominated epoch with $\omega \simeq0$, before their decays finally establish the conventional SM radiation-dominated Universe. Schematically,
\begin{equation}
    \omega_{\rm eff}: \qquad 0\ \longrightarrow\ \frac13\ \longrightarrow\ 0\ \longrightarrow\ \frac13\,.
\end{equation}
This alternating background evolution leads to an unusual temperature history. After a short initial rise, the SM temperature scales as
\begin{equation}
    T(a)\propto
    \begin{dcases}
        a^0 & \text{for nonrelativistic-inflaton domination},\\
        a^{-1/4} & \text{for relativistic-RHN domination},\\
        a^{-3/8} & \text{for nonrelativistic-RHN domination},\\
        a^{-1}   & \text{for SM radiation domination}.
    \end{dcases}
\end{equation}
These scalings differ markedly from those obtained by treating the RHNs as nonrelativistic from production and can have important consequences for any process sensitive to the pre-BBN temperature evolution.

As an application, we study dark-matter (DM) $\chi$ production in this background. First, we consider DM produced directly through a subdominant inflaton decay channel. Although the DM is generated at an early stage, its final abundance is diluted by the entropy released in the later RHN decays. In the hierarchical limit $m_\phi \gg 2\, m_N$, the final yield is controlled by the RHN mass rather than by the inflaton mass, and is parametrically enhanced relative to the conventional reheating result by a factor of order
\begin{equation}
    \frac{Y_\chi^{\rm seesaw}}{Y_\chi^{\rm standard}}
    \sim \frac{m_\phi}{2\, m_N}\,,
\end{equation}
at fixed reheating temperature and inflaton branching fraction. Second, we analyze ultraviolet freeze-in from a reaction density of the form~\cite{McDonald:2001vt, Choi:2005vq, Kusenko:2006rh, Petraki:2007gq, Hall:2009bx, Elahi:2014fsa, Bernal:2017kxu}
\begin{equation}
    \gamma_\chi(T)\propto
    \frac{T^p}{\Lambda^{p-4}}\,.
\end{equation}
The nonstandard temperature and Hubble scalings introduce two critical powers, $p=12$, and $p=20$, which determine whether production in the nonrelativistic-RHN, and relativistic-RHN eras is dominated by the initial or final boundary of each epoch. The intermediate RHN-dominated stages can consequently provide substantial power-law enhancements of the freeze-in abundance.

The remainder of this paper is organized as follows. In Section~\ref{sec:model}, we introduce the inflaton coupling to RHNs in the type-I seesaw model and relate the RHN decay widths to the active-neutrino parameters. In Section~\ref{sec:framework}, we derive the Boltzmann system governing the inflaton, the nonthermal RHN distribution, and the SM radiation bath. In Section~\ref{sec:anal}, we present the numerical evolution and derive analytic approximations for the different cosmological regimes and for the corresponding temperature history. In Section~\ref{sec:DM}, we investigate the consequences for dark-matter production through direct inflaton decays and ultraviolet freeze-in. We summarize our results in Section~\ref{sec:concl}.

%%%%%%%%%%%%%%%%%%%%%%%%%%%%%%%%%%%%%%%%%%%%%%%%%%%%%%%%%%
\section{The model} \label{sec:model}
%%%%%%%%%%%%%%%%%%%%%%%%%%%%%%%%%%%%%%%%%%%%%%%%%%%%%%%%%%
After cosmic inflation, we consider the cosmic energy density to be dominated by a real scalar singlet $\phi$. While this field can be an inflaton which is responsible for inflation, or a curvaton which is responsible for generating the curvature perturbation, or even a reheaton which is responsible for reheating the Universe, we will denote it simply as inflaton, and assume that it oscillated around $\langle\phi\rangle = 0$. At the renormalizable level, $\phi$ can couple to SM Higgs or RHNs $N_i$ $(i=1,2,...,n)$ in the type-I seesaw model~\cite{Minkowski:1977sc, Yanagida:1979as, Glashow:1979nm, Gell-Mann:1979vob, Mohapatra:1979ia}. We assume that the RHN portal dominates and consider the following Lagrangian
\begin{equation}\label{eq:lag}
    -{\cal L} \supset \frac12\, m_\phi^2\, \phi^2 + \left[\frac12 \left( M_i + \lambda_i\, \phi\right) \overline{N_i}\, N_i^c + y_{i\alpha}\, \epsilon_{ab}\, \overline{N_i}\, \ell_\alpha^a\, H^b + \textrm{H.c.}\right],
\end{equation}
where $H$ and $\ell_{\alpha}$ are, respectively, the $SU(2)_L$ Higgs and lepton doublets with $\alpha=e,\mu,\tau$ the lepton flavor index, $\epsilon_{ab}$ $(a,b=1,2)$ is the $SU(2)_{L}$ antisymmetric tensor with $\epsilon_{12}=1=-\epsilon_{21}$ and $y_{i\alpha}$ is a general complex Yukawa coupling. Without loss of generality, we will work on a basis in which the charged lepton Yukawa and the Majorana mass matrix are diagonal and assume that the inflaton coupling is aligned with the RHN mass matrix.

The interaction term between $\phi$ and $N_i$ in Eq.~\eqref{eq:lag} induces the decays of $\phi$ to RHNs, whose total width into RHNs is\footnote{We work in the perturbative regime in which the inflaton-induced contribution to the RHN mass and nonadiabatic fermion production can be neglected; a complete investigation of the complementary regime is left for future work.}
\begin{equation}
    \Gp = \sum_i^{n^\prime}\frac{|\lambda_i|^2}{16\pi}\, m_\phi \left[1 - \left(\frac{2\,M_{i}}{m_\phi}\right)^2\right]^{3/2} ,
\end{equation}
with $n^\prime$ referring to the $N_i$ fulfilling $M_i < m_\phi/2$. On the other hand, the interactions controlled by the Yukawa couplings lead to the decay of $N_i$ into a lepton and a Higgs boson. The total decay width can be expressed as 
\begin{equation}
    \Gamma_{N_i} = \frac{\left(y\, y^\dagger\right)_{ii}\, M_i}{8\pi}\,,
\end{equation}
in the heavy-RHN, zero-temperature approximation, neglecting final-state masses and finite-temperature corrections.

In this model, light-neutrino masses arise through the seesaw mechanism~\cite{Minkowski:1977sc, Yanagida:1979as, Glashow:1979nm, Gell-Mann:1979vob, Mohapatra:1979ia}, with the light neutrino mass matrix given by
\begin{equation}
    m_{\nu} = -v^2\, y^T\, M^{-1}\, y\,,
\end{equation}
where $M = {\rm diag}(M_1,M_2,...,M_n)$, $v \equiv \langle H\rangle \simeq 174$~GeV is the Higgs vacuum expectation value. In terms of observables, we have $m_\nu = U^*\, \hat{m}\, U^\dagger$ where $U$ is the PMNS matrix and $\hat{m} = \textrm{diag}\left(m_1, m_2, m_3\right)$ are the masses of light neutrinos. Experimentally, we have determined $\left|m_3^2 - m_1^2\right| \equiv \left|\Delta m_{\textrm{atm}}^2\right| \simeq 2.5 \times 10^{-3}\,\textrm{eV}^2$ and $m_2^2 - m_1^2 \equiv \Delta m_{\textrm{sol}}^2 \simeq 7.4 \times 10^{-5}\,\textrm{eV}^2$~\cite{Esteban:2024eli, deSalas:2020pgw}.

Using the Casas-Ibarra parametrization~\cite{Casas:2001sr}
\begin{equation}
    y = \frac{i}{v}\, \sqrt{M}\, R\, \sqrt{\hat{m}}\, U^{\dagger}\,,
\end{equation}
where $R$ is a complex orthogonal matrix $R\, R^T = I_{n\times n}$ and $R^T\, R = \tilde{I}_{3\times3}$ where for $n \leq 3$, there will be $3-n$ zero entries on the diagonal of $\tilde{I}_{3\times3}$, we can rewrite
\begin{equation}
    \Gamma_{N_i} = \frac{\left(R\, \hat{m}\, R^{\dagger}\right)_{ii}\, M_i^2}{8 \pi v^2} = \frac{M_i^2}{8 \pi\,  v^2}\, \sum_j m_j \left|R_{ij}\right|^2\, \,.
\end{equation}
Although $\Gamma_{N_i}$ is sensitive to $\hat{m}$, it does not depend on $U$. To explain the neutrino oscillation phenomena with at least two nonzero masses, we need $n \geq 2$. 

For $n = 2$, the lightest active neutrino is massless at tree level: $m_1 = 0$ for normal mass ordering (NO) or $m_3 = 0$ for inverse mass ordering (IO). In either case, we have
\begin{equation}
    \Gamma_{N_i} = \frac{M_i^2}{8 \pi\, v^2} \left(m_2\left|R_{i2}\right|^{2}+m_{h}\left|R_{ih}\right|^{2}\right),
\end{equation}
where $m_h = m_3 = \sqrt{\left|\Delta m_{\textrm{atm}}^{2}\right|}$, $m_2=\sqrt{\Delta m_{\textrm{sol}}^{2}}$ for NO and $m_{h}=m_1=\sqrt{\left|\Delta m_{\textrm{atm}}^{2}\right|}$, $m_2=\sqrt{\Delta m_{\textrm{sol}}^{2}+\left|\Delta m_{\textrm{atm}}^{2}\right|}$ for IO. From the properties of $R$, we obtain the lower bounds
\begin{equation} \label{eq:GNmin}
    \Gamma_{N_i} \geq \frac{M_i^2}{8\pi v^2}
    \begin{dcases} 
        \sqrt{\Delta m_{\textrm{sol}}^{2} }&\textrm{for NO,} \\
        \sqrt{\left|\Delta m_{\textrm{atm}}^{2}\right|} &\textrm{for IO.}
    \end{dcases}
\end{equation}
Alternatively, for $n=3$, we have a 
%potentially 
weaker lower bound
\begin{equation}
    \Gamma_{N_i} \geq \frac{M_i^2}{8\pi\, v^2}\, m_0\,,
    \label{eq:lower_bound_n3}
\end{equation}
where $m_0$ is the lightest neutrino mass scale.

A few comments on leptogenesis~\cite{Fukugita:1986hr, Davidson:2008bu, Fong:2012buy}. Leptogenesis works for $n\geq 2$. In seesaw cosmology, RHNs dominate the cosmic energy density of the Universe before the completion of reheating. In this case, in comparison to the standard scenario, one can achieve an enhancement of the efficiency up to a factor of about 50 which is obtained from the compensation between the gain from the RHN energy density and the loss from entropy dilution, since the cosmic entropy originated from RHNs~\cite{Bernal:2017zvx}. A complete exploration of the leptogenesis parameter space in seesaw cosmology will be left for future work.

Although at least two RHNs are required to reproduce the observed light-neutrino spectrum, we assume that the inflaton decays predominantly into one species $N_k$, with $\Gamma_\phi(\phi \to N_k N_k) \gg \Gamma_\phi(\phi \to N_j N_j)$ for $k \ne j$. In the next section, the cosmological equations therefore track this single nonthermal population, with $m_N = M_k$ and
\begin{align}
    \Gamma_N &\equiv \frac{m_{\rm eff}\, m_N^2}{8\pi v^2}, \\
    \Gamma_\phi &\equiv \frac{|\lambda|^2\, m_\phi}{16\pi} \left(1 - 4\, r_N^2\right)^{3/2},
\end{align}
with $r_N \equiv m_N/m_\phi$ and $m_{\rm eff}$ denotes a light neutrino mass scale encapsulating the lower bounds in Eqs.~\eqref{eq:GNmin} and~\eqref{eq:lower_bound_n3}. The remaining RHNs contribute to the light-neutrino mass matrix but are negligibly populated by inflaton decays.

%%%%%%%%%%%%%%%%%%%%%%%%%%%%%%%%%%%%%
\section{Seesaw cosmology} \label{sec:framework}
%%%%%%%%%%%%%%%%%%%%%%%%%%%%%%%%%%%%%
After cosmic inflation, the total energy density of the Universe is dominated by a nonrelativistic inflaton field $\phi$ coherently-oscillating in a quadratic potential; which defines the cosmic frame. The evolution of the number density $\np$ of inflatons, the number density $\nrh$ of RHNs, and the energy density $\rR$ of SM radiation is governed by the following Boltzmann equations (see, e.g. Refs.~\cite{Buchmuller:2011mw, Coy:2024itg} for derivations in a similar context\footnote{A similar set of Boltzmann equations has been derived in Ref.~\cite{Cosme:2024ndc}; however, a constant equation-of-state parameter for $N$ was assumed, and the time-dilation was ignored. The time-dilation makes a substantial difference for ultrarelativistic RHNs. In addition, Ref.~\cite{Mambrini:2026tla} considered a different model but shares the same Boltzmann equations. They also assumed a constant equation of state, but with an \emph{effective} time-dilate decay width. More importantly, their source term for the SM radiation differs from ours.})
\begin{align}
    \frac{d\np}{dt}  + 3\, H\, \np  &= -   \Gp\, \np\,, \label{eq:BE1}\\
    \frac{d\nrh}{dt} + 3\, H\, \nrh &= +2\,\Gp\, \np - \gamma_N\,,\label{eq:BE2}\\
    \frac{d\rR}{dt}  + 4\, H\, \rR  &= +\Gamma_N\, m_N\, \nrh\,,\label{eq:BE3}
\end{align}
where $H\equiv d\ln a/dt$, with $a$ the scale factor, is the Hubble expansion rate given by the first Friedmann equation
\begin{equation} \label{eq:H}
    H^2 = \frac{\rp + \rrh + \rR}{3\, M_P^2}\,,
\end{equation}
with $M_P \simeq 2.4 \times 10^{18}$~GeV the reduced Planck mass, and $\gamma_N$ is the number-density depletion term defined by
\begin{equation} \label{eq:gammaN}
    \gamma_N(a) \equiv \int_{a_I}^a da'\, \GN(a, a')\, \delta n_N(a,a')\,,
\end{equation}
as a function of the cosmic scale factor $a$, where $a = a_I$ corresponds to the beginning of reheating, where $\GN(a,a')$ is the decay rate of the cohort of $N$ produced at $a'$, measured in the cosmological frame, and $\delta n_N(a,a')$ is the RHN number density distribution (both quantities to be defined next). We notice that, in general, one Boltzmann equation must be solved per RHN; for simplicity, here we focus on the case with a single RHN.

Explicitly, $\GN$ is given by
\begin{equation}
    \GN(a,a') = \frac{m_N}{E_N(a,a')}\, \Gamma_N\,,
\end{equation}
where $E_N$ is the energy of the RHNs in the $\phi$ rest frame that depends on their redshifted momentum $p_N$:
\begin{align}
    E_N^2(a, a') &= m_N^2 + p_N^2(a, a')\,, \label{eq:EN}\\
    p_N^2(a, a') &= \left(\frac{a'}{a}\right)^2 \left(\frac{m_\phi^2}{4}-m_N^2\right).
\end{align} 
$\delta n_N(a, a')\, da'$ corresponds to the number density at epoch $a$ of RHNs produced between $a'$ and $a' + da'$ that have survived until $a$:
\begin{equation}
    \delta n_N(a, a') = \frac{2\, \Gp\, \np(a')}{a'\, H(a')} \left(\frac{a'}{a}\right)^3 \exp\left[-\int_{a'}^a \frac{\GN(a'',a')}{a''\, H(a'')}\, da''\right]
\end{equation}
where the term $2\, \Gp\, \np/(a'\, H)$ corresponds to the production from inflaton decays, the term $(a'/a)^3$ to the dilution, and the exponential factor reflects the decays of the RHNs.

The total RHN number density is
\begin{equation}
    n_N(a) = \int_{a_I}^a \delta n_N(a, a')\, da',
\end{equation}
where the integral accounts for all RHNs produced since $a = a_I$. Several comments: $i)$ For nonrelativistic RHN decays, $\GN \to \Gamma_N$ and, therefore, the standard expression $\gamma_N = \Gamma_N\, n_N$ is recovered. $ii)$ The inflaton decay rate $\Gp$ has no time-dilation factor because the inflaton is taken to be at rest. And $iii)$, the source term for SM radiation comes from
\begin{equation}
    \int_{a_I}^a da'\, E_N(a, a')\, \GN(a, a')\, \delta n_N(a,a') = \Gamma_N\, m_N\, \nrh\,.
\end{equation}

RHNs are continuously generated by inflaton decays, with the same energy $m_\phi/2$ at production. We treat nonthermal RHNs as collisionless between their production and decay, neglecting elastic scattering, number-changing reactions, and inverse inflaton decays. Therefore, their energy is not redistributed between themselves, and hence their inverse decay (that is, the production of inflatons) is always kinematically blocked. Also, we consider small inflaton widths $\Gp$ compared to the Hubble rate, so that the Pauli blocking due to the inflaton decay to fermions is negligible~\cite{Bernal:2026ljj}. In addition, we assume that the SM daughters thermalize on a timescale much shorter than $H^{-1}$, so that their energy density can be assigned a temperature $T$. Because during reheating the energy density from the SM thermal bath is subdominant, the production of RHNs from the thermal bath is always subdominant with respect to the population produced from the inflatons.

The Boltzmann equations~\eqref{eq:BE1} to~\eqref{eq:BE3} are coupled not only through decays but also through the Hubble rate~\eqref{eq:H}, which depends on the sum of all energy densities. As the inflaton is supposed to be nonrelativistic, its energy density $\rp$ is simply
\begin{equation}
    \rp(a)  = m_\phi\, \np(a)\,.
\end{equation}
In turn, the energy density of RHNs is given by
\begin{equation} \label{eq:rhoN}
    \rrh(a) = \int_{a_I}^a \delta n_N(a,a') \times E_N(a, a')\, da'.
\end{equation}
Similarly, their pressure $P_N$ can be computed by integrating $p_N^2/(3\, E_N)$ instead of $E_N$, and is given by
\begin{equation}
    P_N(a) = \int_{a_I}^a \delta n_N(a,a') \times \frac{p_N^2(a,a')}{3\, E_N(a,a')}\, da'.
\end{equation}
Finally, the energy density $\rR$ of SM radiation is given by
\begin{equation}
    \rho_R(a)  = \frac{\Gamma_N\, m_N}{a^4} \int_{a_I}^a \frac{n_N(a')\, {a'}^3}{H(a')} \,da',
\end{equation}
and can be related to the SM temperature $T$ through the expression
\begin{equation}
    \rR(T) = \frac{\pi^2}{30}\, \gs\, T^4,
\end{equation}
with $\gs(T)$ the number of relativistic degrees of freedom.

As a last step, the effective equation-of-state parameter $\omega_{\rm eff}(a)$ of the Universe is defined as
\begin{equation}
    \omega_{\rm eff}(a) \equiv \frac{\sum _i P_i}{\sum_i \rho_i} \simeq \frac{P_N + \frac13\, \rho_R}{\rho_\phi + \rho_N + \rho_R}\,,
\end{equation}
taking into account that the inflaton is pressureless as it is nonrelativistic, and that the SM energy density is dominated by relativistic particles. We emphasize that the treatment of the number-density depletion term in Eq.~\eqref{eq:gammaN} allows to track the smooth transition of the RHN from ultrarelativistic to nonrelativistic, an aspect commonly ignored in the literature.

The system of equations previously presented must be evolved from the beginning of reheating at $a = a_I$ assuming that at that moment all the energy density is stored in the nonrelativistic inflaton. Consequently, one has to take $\rR(a_I) = 0$, $\nrh(a_I) = 0$, and $\np(a_I) = 3\, M_P^2\, H_I^2 / m_\phi$, with $m_\phi$ being the inflaton mass and $H_I \equiv H(a_I)$ the Hubble rate at the end of inflation and the beginning of reheating. We emphasize that the transition of the RHN energy density from relativistic to nonrelativistic makes the system integro-differential. Even if this system has to be solved numerically, we also present analytic estimates in the following.

%%%%%%%%%%%%%%%%%%%%%%%%%%%%%%%%
\section{Analytic solutions} \label{sec:anal}
%%%%%%%%%%%%%%%%%%%%%%%%%%%%%%%%
Approximate analytic solutions can be obtained by dividing the evolution into four asymptotic regimes and neglecting the subdominant energy densities and decay depletion within each regime. They correspond to $i)$ inflaton domination ($a_I \lesssim a \lesssim a_\phi$), $ii)$ relativistic RHN domination ($a_\phi \lesssim a \lesssim \anr$), $iii)$ nonrelativistic RHN domination ($\anr \lesssim a \lesssim \arh$), and $iv)$ SM radiation domination ($\arh \lesssim a$). A well-separated four-stage evolution requires approximately $H_I \gg \Gp$, $m_\phi \gg 2\, m_N$, and $\Gp \gg \Gamma_N$. The four regimes are presented below.

%%%%%%%%%%%%%%%%%%%%%%%%%%%%%%%%
\subsection{Inflaton domination}
%%%%%%%%%%%%%%%%%%%%%%%%%%%%%%%%
After the end of cosmic inflation, the energy density of the Universe is dominated by nonrelativistic inflatons (that is, $H \propto \sqrt{\rp}$ with $\rp(a) \propto a^{-3}$) that decay into a pair of longlived RHNs. Their corresponding number densities are
\begin{align}
    \np(a) &\simeq 3\, \frac{M_P^2\, H_I^2}{m_\phi} \left(\frac{a_I}{a}\right)^3, \label{eq:nphi1}\\
    \nrh(a) &\simeq 4\, \frac{M_P^2\, H_I\, \Gp}{m_\phi} \left(\frac{a_I}{a}\right)^\frac32 \left[1 - \left(\frac{a_I}{a}\right)^\frac32\right]. \label{eq:nN1}
\end{align}
In turn, the energy densities are
\begin{align}
    \rp(a) &\simeq 3\, M_P^2\, H_I^2 \left(\frac{a_I}{a}\right)^3,\\
    \rrh(a) &\simeq \frac65\, M_P^2\, H_I\, \Gp \left(\frac{a_I}{a}\right)^\frac32 \left[1 - \left(\frac{a_I}{a}\right)^\frac52\right],\\
    \rR(a) &\simeq M_P^2\, \Gp\, \Gamma_N\, \frac{m_N}{m_\phi} \left[1 - \frac85 \left(\frac{a_I}{a}\right)^\frac32 + \frac35 \left(\frac{a_I}{a}\right)^4\right]. \label{eq:rR1}
\end{align}
We note that, even if relativistic, the RHN energy density does not scale as free radiation (that is, $\rho(a) \propto a^{-4}$) due to the source term. In addition, the energy density of SM rapidly reaches a plateau as it comes from the decay of a relativistic particle (i.e., the RHN) sourced in turn by the decay of a nonrelativistic particle (i.e., the inflaton)~\cite{Cosme:2024ndc}.

The end of the inflaton dominated era is defined as the moment at which $\rp(\aph) = \rrh(\aph)$, and corresponds to a scale factor $\aph$ given by
\begin{equation}\label{eq:aphi}
    \aph \simeq a_I \left(\frac52\, \frac{H_I}{\Gp}\right)^\frac23.
\end{equation}

%%%%%%%%%%%%%%%%%%%%%%%%%%%%%%%%
\subsection{Relativistic RHN domination}
%%%%%%%%%%%%%%%%%%%%%%%%%%%%%%%%
After the end of the inflaton dominated era, inflaton number density is exponentially suppressed, and the Hubble expansion is dominated by relativistic RHNs. The energy densities of $N$ and the SM radiation are
\begin{align}
    \rrh(a) &\simeq \frac{12}{25}\, M_P^2\, \Gp^2 \left(\frac{\aph}{a}\right)^4,\\
    \rR(a) &\simeq \frac43\, M_P^2\, \Gp\, \Gamma_N\, \frac{m_N}{m_\phi}\, \frac{a_\phi}{a} \left[1 - \frac14 \left(\frac{a_\phi}{a}\right)^3\right]. \label{eq:rR2}
\end{align}
We note that, due to the source term, $\rR(a) \propto a^{-1}$.

The transition from relativistic to nonrelativistic can be estimated by computing the moment at which the momenta of all RHNs are smaller than $m_N$. Taking into account that the highest momentum corresponds to particles produced at $a = \aph$, the scale factor $\anr$ at the transition is given by
\begin{equation}\label{eq:anr}
    \anr = \aph\, \sqrt{\left(\frac{m_\phi}{2\, m_N}\right)^2 - 1}\,.
\end{equation}

%%%%%%%%%%%%%%%%%%%%%%%%%%%%%%%%
\subsection{Nonrelativistic RHN domination}
%%%%%%%%%%%%%%%%%%%%%%%%%%%%%%%%
In this era, the expansion of the Universe is still dominated by RHNs, but this time they are nonrelativistic. The energy densities are
\begin{align}
    \rrh(a) &\simeq \frac{12}{25}\, M_P^2\, \Gp^2 \left[\left(\frac{m_\phi}{2\, m_N}\right)^2 - 1\right]^{-2} \left(\frac{\anr}{a}\right)^3,\\
    \rR(a) &\simeq \frac65\, M_P^2\, H_I\, \Gamma_N \left[\left(\frac{m_\phi}{2\, m_N}\right)^2 - 1\right]^{-\frac14} \left(\frac{a_I}{a}\right)^\frac32. \label{eq:rR3}
\end{align}
As expected, this era corresponds to a standard early matter domination, in which $\rrh(a) \propto a^{-3}$ and $\rR(a) \propto a^{-3/2}$. In addition, at late times, the SM energy density depends on $\Gamma_N$ and not on $\Gp$ as the dynamics is IR dominated; that is, it is independent of how the RHNs were produced.

The end of the reheating era, defined as the onset of SM domination, can be estimated by the equality of the energy densities of $N$ and SM radiation, and corresponds to a scale factor
\begin{equation}\label{eq:arh}
    \arh \simeq a_I \left[\frac{25}{2}\, \frac{H_I^2\, m_N}{\Gamma_N^2\, \sqrt{m_\phi^2 - 4\, m_N^2}}\right]^\frac13.
\end{equation}

%%%%%%%%%%%%%%%%%%%%%%%%%%%%%%%%
\subsection{SM radiation domination}
%%%%%%%%%%%%%%%%%%%%%%%%%%%%%%%%
After the end of the reheating era, the abundance of RHNs is exponentially suppressed, and the Universe is dominated by SM radiation with an energy density
\begin{equation}
    \rR(a) \simeq \frac{12}{25}\, M_P^2\, \Gamma_N^2 \left(\frac{\arh}{a}\right)^4, \label{eq:rR4}
\end{equation}
that scales as free radiation $\rR(a) \propto a^{-4}$ and, again, depends on $\Gamma_N$ but not on $\Gp$.

%%%%%%%%%%%%%%%%%%%%%%%%%%%%%%%%
\subsection{Evolution of the SM temperature}
%%%%%%%%%%%%%%%%%%%%%%%%%%%%%%%%
From the evolution of the SM energy density $\rR$ in Eqs.~\eqref{eq:rR1}, \eqref{eq:rR2}, \eqref{eq:rR3} and~\eqref{eq:rR4}, the SM temperature can be derived and given by\footnote{We note that Ref.~\cite{Cosme:2024ndc} reported a scaling $T(a) \propto a^{-1/2}$ for the case of the decay of a ultrarelativistic RHN (that is, $\aph \leq a \leq \anr$). The difference is due to the absence of the time-dilation in the RHN decay in their analysis.}
\begin{equation}
    T(a) \simeq
    \begin{dcases}
        \Tmax &\text{for } a_I \leq a \leq \aph\,,\\
        \Tmax  \left(\frac{\aph}{a}\right)^\frac14 &\text{for } \aph \leq a \leq \anr\,,\\
        \Trh \left(\frac{\arh}{a}\right)^\frac38 &\text{for } \anr \leq a \leq \arh\,,\\
        \Trh \left(\frac{\arh}{a}\right) &\text{for } \arh \leq a\,.
    \end{dcases}
\end{equation}
Here, $\Tmax$ is the maximum temperature reached by the thermal plasma, defined by
\begin{equation}\label{eq:Tmax}
    \Tmax^4 \simeq \frac{30}{\pi^2\, \gs}\, M_P^2\, \Gp\, \Gamma_N\, \frac{m_N}{m_\phi}\,,
\end{equation}
which rapidly approaches a plateau in the period $a_I \lesssim a \lesssim \aph$, in which the RHN is relativistic and is sourced by the decay of the nonrelativistic inflaton~\cite{Cosme:2024ndc}. The SM temperature starts to decrease only once the RHNs are not sourced anymore, scaling as $T(a) \propto a^{-1/4}$ during the era $\aph \lesssim a \lesssim \anr$. Once the RHNs become nonrelativistic, a standard early matter dominated era takes place with $T(a) \propto a^{-3/8}$~\cite{Giudice:2000ex}. Finally, SM radiation starts to dominate the energy budget of the Universe, at a temperature $\Trh$ given by
\begin{equation} \label{eq:Trh}
    \Trh^2 \simeq \frac{6}{\pi} \sqrt{\frac{2}{5\, \gs}}\, M_P\, \Gamma_N\,.
\end{equation}
As a standard signature of the early matter dominated era, $\Trh \propto \sqrt{M_P\, \Gamma_N}$, only depends on the decay width of the particle that sources the SM radiation, and not on $\Gp$. After the end of reheating, SM radiation scales as free radiation and therefore $T(a) \propto a^{-1}$.

To quantify the era of RHN-domination while nonrelativistic and relativistic, it is convenient to define $R_1 \equiv a_{\rm rh}/a_{\rm nr}$ and $R_2 \equiv a_{\rm nr}/a_{\phi}$, respectively. Using Eqs.~\eqref{eq:aphi}, \eqref{eq:anr} and~\eqref{eq:arh}, we obtain
\begin{align}
    R_1^3 &= \frac{4\, |\lambda|^4\, v^4}{m_{\rm eff}^2\, m_\phi^2} \left(1 - 4\, r_N^2\right),\\
    R_2 &= \frac{1}{2\,r_N}\, \sqrt{1 - 4\, r_N^2}\,,
\end{align}
which do not explicitly depend on $H_I$. While $R_2 \gg 1$ requires $r_N \ll 1$, $R_1 \gg 1$ with $r_N \ll 1$ translates to
\begin{equation}
    m_{\rm eff} \ll 0.06\,{\rm eV}
    \left(\frac{|\lambda|}{10^{-3}}\right)^2
    \left(\frac{10^{9}\,{\rm GeV}}{m_\phi}\right).
\end{equation}

%%%%%%%%%%%%%%%%%%%%%%%%%%%%%%%%%%%%%%%%%%%%%%%%%%%
 \begin{figure}[t!]
     \def\sepf{0.49}
     \centering
     \includegraphics[width=\sepf\columnwidth]{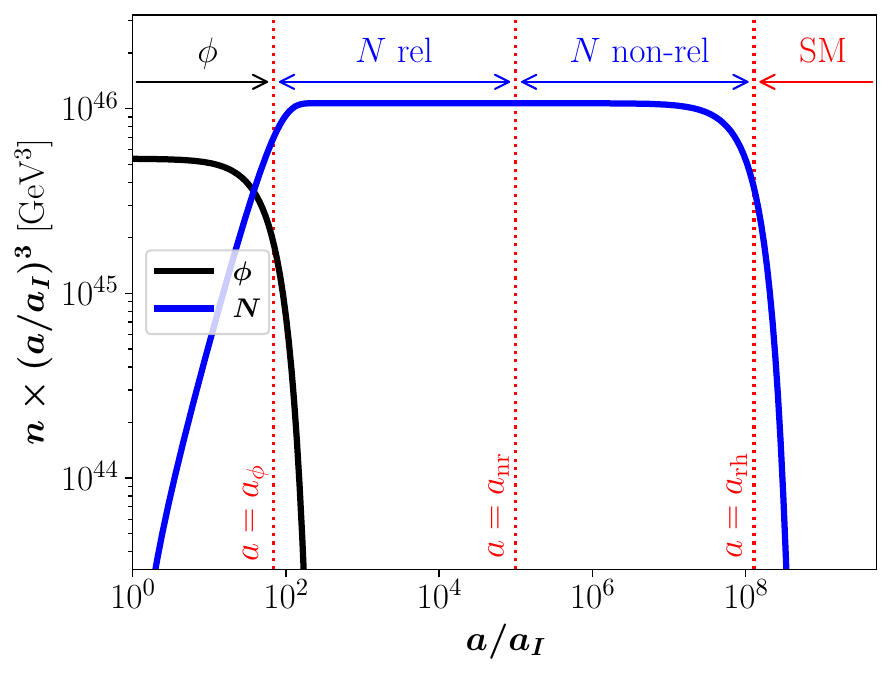}
     \includegraphics[width=\sepf\columnwidth]{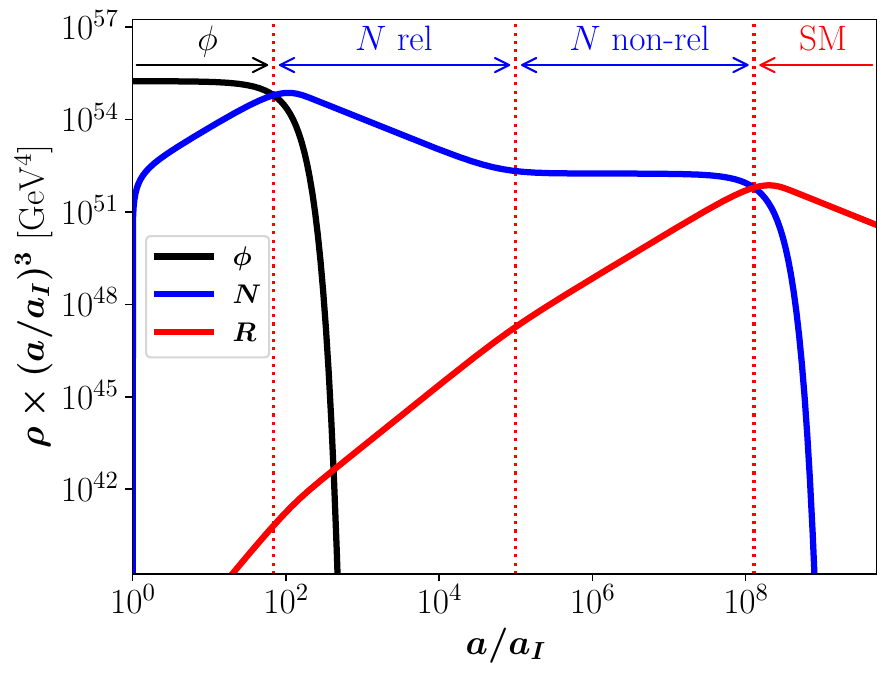}
     \includegraphics[width=\sepf\columnwidth]{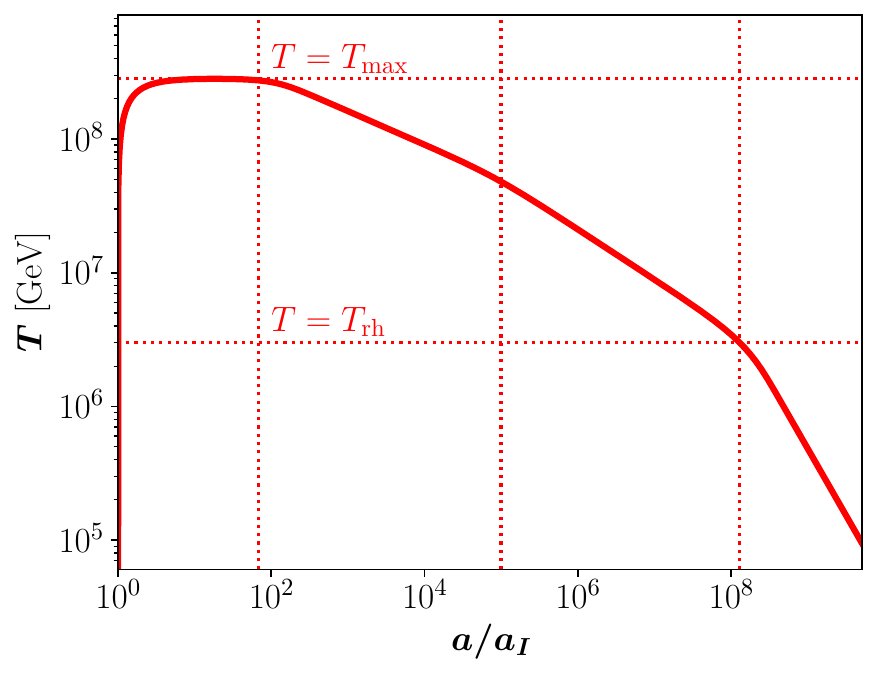}
     \includegraphics[width=\sepf\columnwidth]{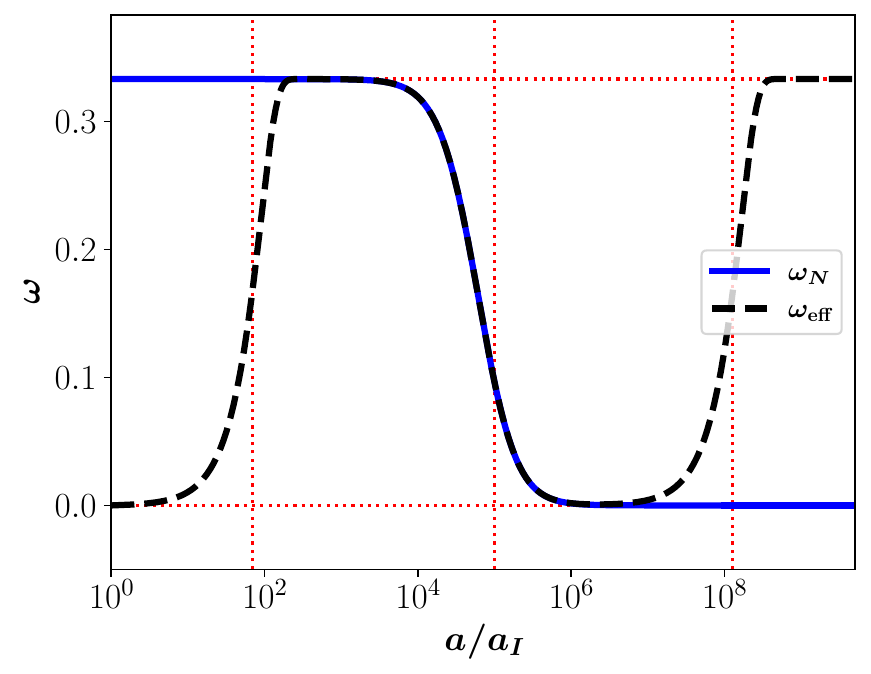}
     \caption{Evolution of the Universe in the cosmological seesaw scenario with $n = 2$ RHNs, for $H_I = 10^9$~GeV, $m_\phi = 3.3 \times 10^9$~GeV, $m_N = 1.7 \times 10^6$~GeV, $\Gp = 2.5 \times 10^6$~GeV and $\Gamma_N = 3.1\times 10^{-5}$~GeV. Solid black, blue, and red lines correspond to quantities related to the inflaton, RHN and SM radiation, respectively. The number densities and the energy densities are shown in the upper panels. The lower panels show the SM temperature (left), and the equations of state of the RHN and the Universe (right).}
     \label{fig:evoA}
 \end{figure} 
%%%%%%%%%%%%%%%%%%%%%%%%%%%%%%%%%%%%%%%%%%%%%%%%%%%
Figure~\ref{fig:evoA} shows an example of the cosmic evolution of the Universe with $n = 2$ RHNs, for $H_I = 10^9$~GeV, $m_\phi = 3.3 \times 10^9$~GeV, $m_N = 1.7 \times 10^6$~GeV and $\Gp = 2.5 \times 10^6$~GeV; the decay width $\Gamma_N = 3.1\times 10^{-5}$~GeV (which implies $m_{\rm eff} \simeq 8.2$~meV) was chosen by minimizing the bound in Eq.~\eqref{eq:GNmin}. This gives $R_1 \simeq 1942$ and $R_2 \simeq 971$. In addition to the constraints of neutrino masses, it was also demanded $m_\phi > H_I$ to allow the inflaton to oscillate at the bottom of its potential, perturbative coupling $\lambda$, a RHN mass $m_N \gg v$ to have a kinematically open RHN decay into a Higgs and a lepton, an inflationary scale $H_I \lesssim 6\times 10^{13}$~GeV in accordance with cosmological observations by BICEP/Keck~\cite{Baumann:2009ds, BICEP:2021xfz}, and a derived reheating temperature $\Trh > T_\text{BBN} \simeq 4$~MeV~\cite{Sarkar:1995dd, Kawasaki:2000en, Hannestad:2004px, Barbieri:2025moq} to avoid spoiling the success of BBN. Furthermore, the parameters have also been chosen to feature long and well-separated asymptotical regimes, as previously described. Solid black, blue, and red lines correspond to quantities related to the inflaton, RHN, and SM radiation, respectively. The number densities and the energy densities are shown in the upper panels. The red dotted vertical lines correspond to $a = \aph$, $a = \anr$, and $a = \arh$, and limit the four asymptotic regimes described previously. The lower left panel shows the SM temperature, and the horizontal lines correspond to $T = \Tmax \simeq 3\times 10^8$~GeV and $T = \Trh \simeq 3\times 10^6$~GeV. Finally, the lower right panel shows in blue the equation of state $\omega_N$ of the RHN, and the effective equation of state of the Universe $\omega_{\rm eff}$ with a black dashed line. For reference, the red horizontal dotted lines show $\omega = 0$ (matter) and $\omega = 1/3$ (radiation). We emphasize the smooth transitions matter -- radiation -- matter -- radiation. A comparison between the numerical results and the approximated analytical expression can be found in Appendix~\ref{sec:app}.

%%%%%%%%%%%%%%%%%%%%%%%%%%%%%%%%%%%%%%%%%%%%%%%%%%%
 \begin{figure}[t!]
     \def\sepf{0.49}
     \centering
     \includegraphics[width=\sepf\columnwidth]{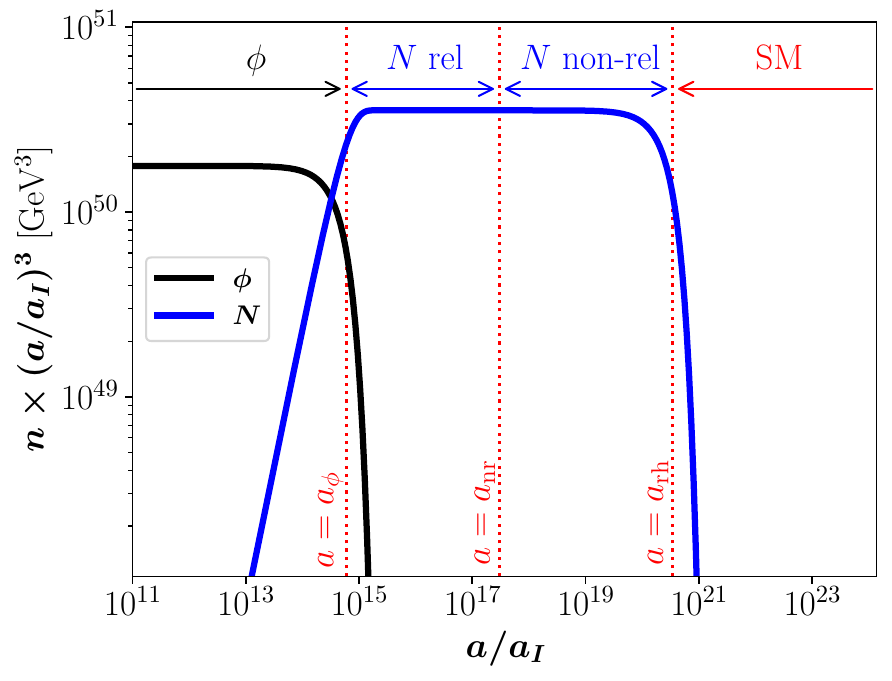}
     \includegraphics[width=\sepf\columnwidth]{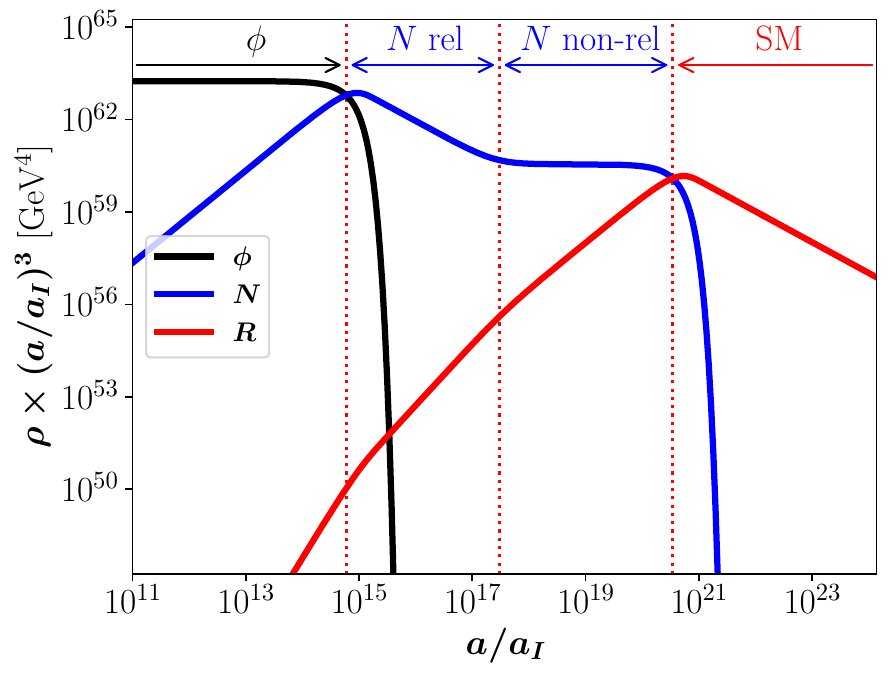}
     \includegraphics[width=\sepf\columnwidth]{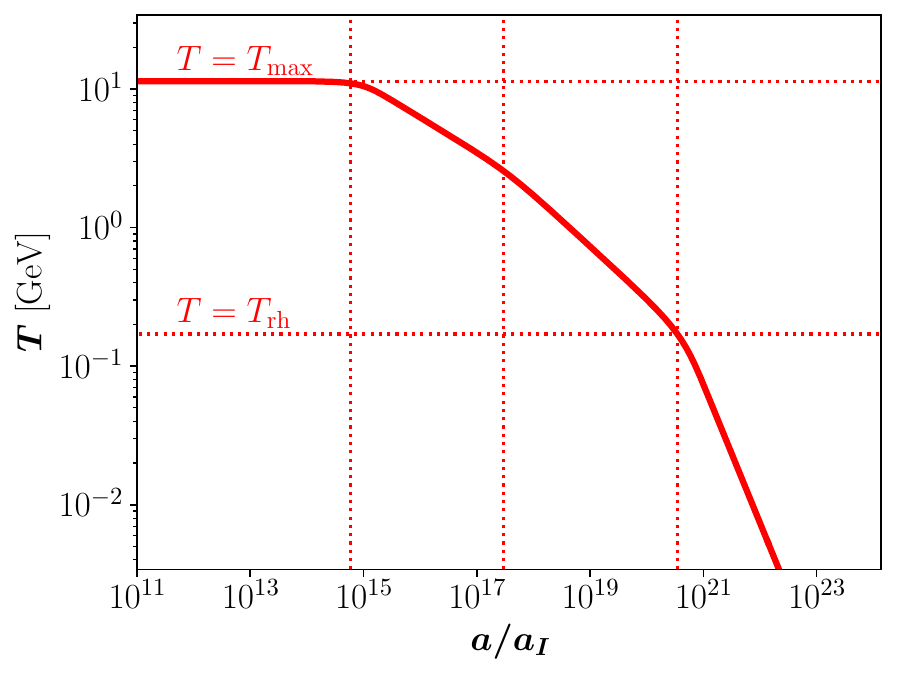}
     \includegraphics[width=\sepf\columnwidth]{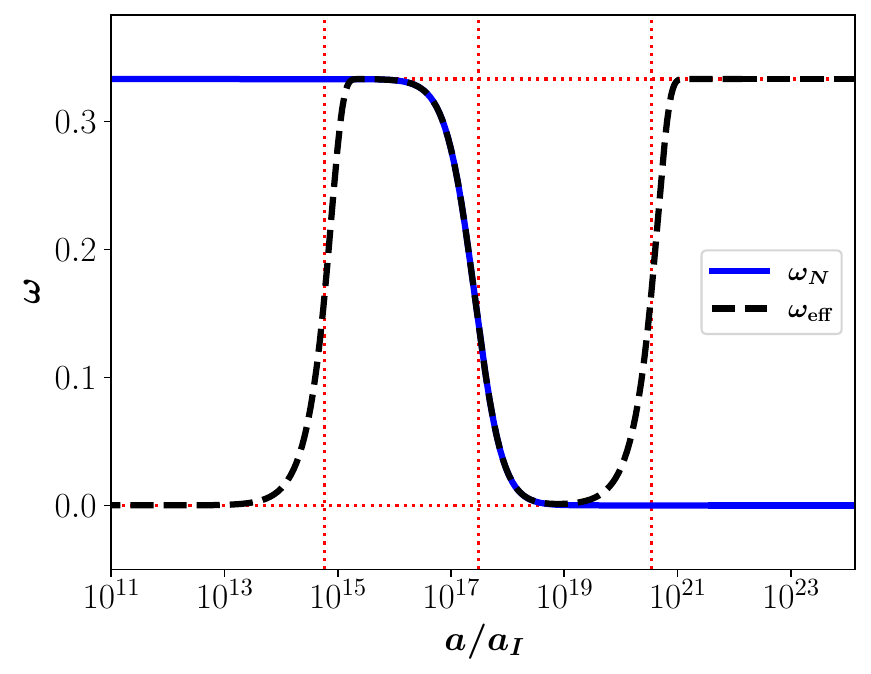}
     \caption{Similar to Fig.~\ref{fig:evoA} but for $n = 3$ RHNs, with $H_I = m_\phi = 10^{13}$~GeV, $m_N = 10^{10}$~GeV, $\Gp = 10^{-9}$~GeV and $\Gamma_N = 10^{-19}$~GeV.}
     \label{fig:evoC}
 \end{figure} 
%%%%%%%%%%%%%%%%%%%%%%%%%%%%%%%%%%%%%%%%%%%%%%%%%%%
A second example of seesaw cosmology is shown in Fig.~\ref{fig:evoC}, corresponding to $n = 3$ RHNs, with $H_I = m_\phi = 10^{13}$~GeV, $m_N = 10^{10}$~GeV, $\Gp = 10^{-9}$~GeV and $\Gamma_N = 10^{-19}$~GeV ($m_{\rm eff} \simeq 7.6 \times 10^{-25}$~eV); which corresponds to $R_1 \simeq 1170$ and $R_2 \simeq 500$. In this case, the decay widths of the RHNs are not fixed by the active-neutrino mass differences but have a weaker lower bound given by Eq.~\eqref{eq:lower_bound_n3}. Even if the inflationary scale $H_I$ is high, the very small decay width of the RHN drags the reheating scale to small temperatures: $\Tmax \simeq 11$~GeV and $\Trh \simeq 0.2$~GeV. 

In all cases, to accurately describe the transitions between difference effective equation of state of the Universe, numerical treatment of the Boltzmann equations using the full distribution of the RHNs is required.

%%%%%%%%%%%%%%%%%%%%%%%%%%%%%%%%%%%%%%%%%%%%%%%%%%%%%%%%%
\section{Dark Matter} \label{sec:DM}
%%%%%%%%%%%%%%%%%%%%%%%%%%%%%%%%%%%%%%%%%%%%%%%%%%%%%%%%%
In this section, the production of DM during the seesaw cosmological scenario is studied. The evolution of the DM number density $\ndm$ obeys the integrated Boltzmann equation
\begin{equation}
    \frac{d\ndm}{dt} + 3\, H\, \ndm = \gamma_\chi\,,
\end{equation}
where $\gamma_\chi$ corresponds to its production rate density, which includes the multiplicity of DM particles produced per reaction. We focus on out-of-equilibrium processes and DM number densities much smaller than the density at equilibrium, so that annihilation terms are always subdominant and are therefore disregarded. In addition, DM is supposed to be stable, or at least to have a lifetime much longer than the age of the Universe, so that decay terms can also be discarded. Assuming a negligible DM population $n_\chi(a_I) = 0$, the DM number density is
\begin{equation} \label{eq:nchi}
    n_\chi(a) = \left(\frac{\arh}{a}\right)^3\, \mathcal{I}(a_I,a)\,,
\end{equation}
with
\begin{equation}
    \mathcal{I}(a_1,a_2) \equiv \frac{1}{\arh^3}\, \int_{a_1}^{a_2} da'\, \frac{{a'}^2\, \gamma_\chi(a')}{H(a')}\,.
\end{equation}

The corresponding DM yield $Y \equiv n_\chi/s$ is defined as the DM number density normalized to the SM entropy density $s$ given by
\begin{equation}
    s(T) = \frac{2\, \pi^2}{45}\, \gss\, T^3,
\end{equation}
where $\gss(T)$ is the number of relativistic degrees of freedom that contribute to the SM entropy. To match the entire observed DM relic density, it is required that its asymptotic value $Y_0$ at low temperatures satisfies
\begin{equation}\label{eq:obs_DM}
    m_\chi\, Y_0 = \frac{\Omega h^2\, \rho_c}{s(T_0)\, h^2} \simeq 4.7 \times 10^{-10}~\text{GeV},
\end{equation}
where $m_\chi$ is the mass of the DM, $s(T_0) \simeq 2.69 \times 10^3$~cm$^{-3}$ is the present entropy density~\cite{ParticleDataGroup:2024cfk}, $\rho_c \simeq 1.05 \times 10^{-5}~h^2$~GeV/cm$^3$ is the critical energy density of the Universe, and $\Omega h^2 \simeq 0.12$ is the observed DM relic abundance~\cite{Planck:2018vyg}.

Now, two DM production channels will be studied: first through direct inflaton decays, then from the SM thermal plasma.

%%%%%%%%%%%%%%%%%%%%%%%%%%%%%%%%%%%%%%%%%%%%%%%%%%%%%%%%%
\subsection{Inflaton decay}
%%%%%%%%%%%%%%%%%%%%%%%%%%%%%%%%%%%%%%%%%%%%%%%%%%%%%%%%%
$\mathcal{N}_\chi$ DM particles $\chi$ can be created directly from inflaton decays, with a branching fraction $\Br \equiv \Gamma(\phi \to \mathcal{N}_\chi\, \chi) / \Gp^{\rm tot} \ll 1$, with $\Gp^{\rm tot} = \Gamma(\phi \to NN) + \Gamma(\phi \to \mathcal{N}_\chi\, \chi) \simeq \Gamma(\phi\to NN)$. For this channel to be open and not kinematically suppressed, we assume $m_\phi \gg \mathcal{N}_\chi\, m_\chi$. In this case, the production rate density is given by
\begin{equation}
    \gamma_\chi = \mathcal{N}_\chi\, \Br\, \Gp^{\rm tot}\, \np\,.
\end{equation}
We note that DM $\chi$ can be identified with a RHN in the case of $n\geq 3$ where one of them is so feebly coupled to $\ell$ and $H$ that it has a lifetime much longer than the cosmic age.

After all inflatons have decayed, the DM number density at $a \gg \aph$ is given by
\begin{equation} \label{eq:ndm}
    \ndm(a) = \mathcal{N}_\chi\, \Br\, \np(a_I) \left(\frac{a_I}{a}\right)^3 = 3\, \mathcal{N}_\chi\, \Br\, \frac{M_P^2\, H_I^2}{m_\phi} \left(\frac{a_I}{a}\right)^3,
\end{equation}
and its corresponding yield at present is, therefore,
\begin{equation} \label{eq:Ydm}
    Y_0 = \frac{n_\chi(\arh)}{s(\Trh)} \simeq \frac{6}{25}\, \mathcal{N}_\chi\, \Br\, \frac{M_P^2}{s(\Trh)}\, \frac{\Gamma_N^2}{m_N} \simeq \frac38\, \mathcal{N}_\chi\, {\rm Br}\, \frac{\gs}{\gss}\, \frac{\Trh}{m_N}\,,
\end{equation}
where in the last step Eq.~\eqref{eq:Trh} was used. It is worth noting that even if the DM number density produced from inflaton decays depends on inflaton properties (decay width through the branching ratio and mass $m_\phi$; cf. Eq.~\eqref{eq:ndm}), once the expansion of the Universe and the entropy dilution are taken into account, the DM yield at present depends on RHN properties (decay width $\Gamma_N$ through $\Trh$ and mass $m_N$; cf. Eq.~\eqref{eq:Ydm}).\footnote{Notice here that, in this scenario a detail treatment of the reheating evolution is not required, in contrast to the scenario addressed in the next section}. In addition, the DM yield at present has exactly the same parametric dependence as the standard case where DM is directly produced from the decay of a nonrelativistic inflaton
\begin{equation}
    Y_0^{\rm standard} = \frac34\, \mathcal{N}_\chi\, {\rm Br}\, \frac{\gs}{\gss}\, \frac{\Trh}{m_\phi}\,;
\end{equation}
see, e.g. Ref.~\cite{Bernal:2021qrl}. Therefore, within seesaw cosmology, the DM production from inflaton decays is enhanced by a factor
\begin{equation}
    \frac{Y_0}{Y_0^{\rm standard}} \simeq \frac{m_\phi}{2\, m_N}\,,
\end{equation}
with $m_\phi \gg 2\, m_N$.

%%%%%%%%%%%%%%%%%%%%%%%%%%%%%%%%%%%%%%%%%%%%%%%%%%%%%%%%%
\subsection{UV freeze-in}
%%%%%%%%%%%%%%%%%%%%%%%%%%%%%%%%%%%%%%%%%%%%%%%%%%%%%%%%%
Alternatively, if DM is connected to SM particles through nonrenormalizable operators with mass dimension $D \geq 5$, DM is produced from the SM thermal bath via UV freeze-in. In addition, if the DM mass is smaller than the reheating temperature ($m_\chi < \Trh$), the interaction rate density can be parametrized as\footnote{We note that Ref.~\cite{Cosme:2024ndc} focuses on the ``freeze-in at stronger coupling'' scenario with $m_\chi \gg \Tmax$, where DM production experiences a strong Boltzmann suppression.}
\begin{equation} \label{eq:UVFI}
    \gamma_\chi(T) = \frac{T^p}{\Lambda^{p-4}}\,,
\end{equation}
where $\Lambda$ corresponds to the mass scale of new physics (and contains couplings, multiplicities, and symmetry factors), and $p = 2\, (D-2) \geq 6$ is the scaling of the rate density for relativistic scattering, as a function of the  SM temperature. The consistency of this effective approach requires $\Tmax < \Lambda$. A particularly well motivated example of this kind of interaction corresponds to the case in which DM is produced in annihilations of SM particles from the thermal bath, mediated by the $s$-channel exchange of gravitons, where $p = 8$ and $\Lambda \simeq M_P$~\cite{Garny:2015sjg, Tang:2017hvq, Garny:2017kha, Bernal:2018qlk, Barman:2021ugy}.

Given this interaction in Eq.~\eqref{eq:UVFI}, DM production can be analytically estimated by integrating Eq.~\eqref{eq:nchi} in the four asymptotic regimes described in Section~\ref{sec:anal}. First, UV freeze-in production during the standard SM radiation dominated era can be computed integrating in the interval $\arh \leq a \leq a_0$, where $a_0$ is the scale factor at present:
\begin{equation}
    \mathcal{I}(\arh,a_0) \simeq \frac{\mathcal{F}}{p-5} \left[1 - \left(\frac{\arh}{a_0}\right)^{p-5}\right] \simeq \frac{\mathcal{F}}{p-5}\,,
\end{equation}
where
\begin{equation}
    \mathcal{F} \equiv \frac{\Trh^p\, \Lambda^{4-p}}{H(\Trh)}\,,
\end{equation}
and
\begin{equation}
    H(\Trh) = \frac{\pi}{3}\, \sqrt{\frac{\gs(\Trh)}{10}}\, \frac{\Trh^2}{M_P}\,.
\end{equation}
During this period, the majority of the DM is produced near $T = \Trh$, as a signature of UV freeze-in.

During the era dominated by nonrelativistic RHNs, that is, $\anr \leq a \leq \arh$,
\begin{equation}
    \mathcal{I}(\anr,\arh) \simeq \mathcal{F} \times
    \begin{dcases}
        \frac{8}{3(12-p)} &\text{for } p < 12\,,\\
        \ln(R_1) &\text{for } p = 12\,,\\
        \frac{8}{3(p-12)}\,  R_1^\frac{3(p-12)}{8} &\text{for } p > 12\,,
    \end{dcases}
\end{equation}
where
\begin{equation}
    R_1 \equiv \frac{\arh}{\anr} > 1\,.
\end{equation}
For $p < 12$, DM is mainly produced at $a = \arh$ with an amount comparable to that of the SM radiation dominated era. However, for $p > 12$, production occurs near $a = \anr$ and is boosted by a factor $R_1^{3(p-12)/8}$. An intermediate case occurs for $p = 12$, in which the enhancement comes only from the logarithm $\ln(R_1)$.

Third, during the era dominated by relativistic RHNs, that is, $\aph \leq a \leq \anr$,
\begin{equation}
    \mathcal{I}(\aph,\anr) \simeq \mathcal{F} \times
    \begin{dcases}
        \frac{4}{20-p}\, R_1^\frac{3(p-12)}{8} &\text{for } p < 20\,,\\
        R_1^3\, \ln(R_2) &\text{for } p = 20\,,\\
        \frac{4}{p-20}\, R_1^\frac{3(p-12)}{8}\, R_2^\frac{p-20}{4} &\text{for } p > 20\,,
    \end{dcases}
\end{equation}
where
\begin{equation}
    R_2 \equiv \frac{\anr}{\aph} > 1\,.
\end{equation}
For $p < 20$, most of the DM is produced near $a = \anr$. However, for $p < 12$, $p = 12$ and $12 < p < 20$, there is a suppression, an equivalent contribution (that is, there is no power-law suppression or enhancement in $R_1$ or $R_2$), or an enhancement with respect to what is produced after reheating, respectively. For $p > 20$, DM is mostly produced near $a = \aph$, with a double power-law enhancement $R_1^\frac{3(p-12)}{8}\, R_2^\frac{p-20}{4}$. Finally, for $p = 20$, the boost is $R_1^3\,  \ln(R_2)$.

Finally, during the inflaton dominated era, that is, $a_I \leq a \leq \aph$,
\begin{equation}
    \mathcal{I}(a_I,\aph) \simeq \frac29\, \mathcal{F}\, R_1^\frac{3(p-12)}{8}\,  R_2^\frac{p-20}{4} \left[1 - \left(\frac{a_I}{\aph}\right)^\frac92\right] \simeq \frac29\, \mathcal{F}\, R_1^\frac{3(p-12)}{8}\,  R_2^\frac{p-20}{4}.
\end{equation}
In this case, for $p \leq 12$ and $p \geq 20$, production is always suppressed and enhanced, respectively. In contrast, for the range $12 < p < 20$ the production could be enhanced or suppressed, depending on the ratios $R_1$ and $R_2$.

%%%%%%%%%%%%%%%%%%%%%%%%%%%%%%%%%%%%%%%%%%%%
\bgroup
\def\arraystretch{1.8}
\begin{table}[t!]
    \centering
    \begin{tabular}{|c|c||c|}
        \hline
        \bf{Dimension} & $\boldsymbol{p}$ & $\boldsymbol{\mathcal{I}(a_I,a_0) / \mathcal{F}}$ \\
        \hline\hline
         $5 \leq D \leq 7$ & $6 \leq p \leq 10$ & $\frac{1}{p - 5} + \frac{8}{3(12 - p)}$ \\
         $D = 8$ & $p = 12$ & $\frac{1}{p - 5} + \frac{4}{20-p} + \ln(R_1)$\\
         $9 \leq D \leq 11$ & $14 \leq p \leq 18$ & $\frac{1}{p - 5} + \left[\frac{8}{3(p-12)} + \frac{4}{20-p} + \frac29\, R_2^{-\frac{20-p}{4}}\right] R_1^\frac{3(p-12)}{8}$\\
         $D = 12$ & $p = 20$ & $\frac{1}{p - 5} + R_1^3\, \ln(R_2) + \left[\frac{8}{3(p-12)} + \frac29\right] R_1^\frac{3(p-12)}{8}$\\
         $D \geq 13$ & $p \geq 22$ & $\frac{1}{p - 5} + \left[\frac{8}{3(p-12)} + \left(\frac29 + \frac{4}{p-20}\right) R_2^\frac{p-20}{4}\right] R_1^\frac{3(p-12)}{8}$\\
         \hline
    \end{tabular}
    \caption{Main contributions to $\mathcal{I}(a_I,a_0)$, as a function of the exponent $p$ or, equivalently, the dimension $D$ of the corresponding operator.}
    \label{tab:I}
\end{table}
\egroup
%%%%%%%%%%%%%%%%%%%%%%%%%%%%%%%%%%%%%%%%%%%%
Once all the regimes are taken into account, the final DM yield at present is given by
\begin{equation}
    Y_0 = \frac{\mathcal{I}(a_I,a_0)}{s(\Trh)} = \frac{135}{2 \pi^3\, \gss(\Trh)}\, \sqrt{\frac{10}{\gs(\Trh)}}\, \frac{M_P\, \Trh^{p-5}}{\Lambda^{p-4}}\, \frac{\mathcal{I}(a_I,a_0)}{\mathcal{F}}\,,
\end{equation}
where the main contributions to $\mathcal{I}(a_I,a_0) = \mathcal{I}(a_I,\aph) + \mathcal{I}(\aph,\anr) + \mathcal{I}(\anr,\arh) + \mathcal{I}(\arh,a_0)$ are summarized in Table~\ref{tab:I}, as a function of the exponent $p$ or, equivalently, the dimension $D$ of the corresponding operator. We note that the previous analytical expressions are only reliable in the limits where $R_1 \gg 1$ and $R_2 \gg 1$. In the general case, the Boltzmann equation for the DM number density has to be solved numerically with the cosmological seesaw background.

%%%%%%%%%%%%%%%%%%%%%%%%%%%%%%%%%%%%%%%%%%%%%%%%%%%
 \begin{figure}[t!]
     \def\sepf{0.49}
     \centering
     \includegraphics[width=\sepf\columnwidth]{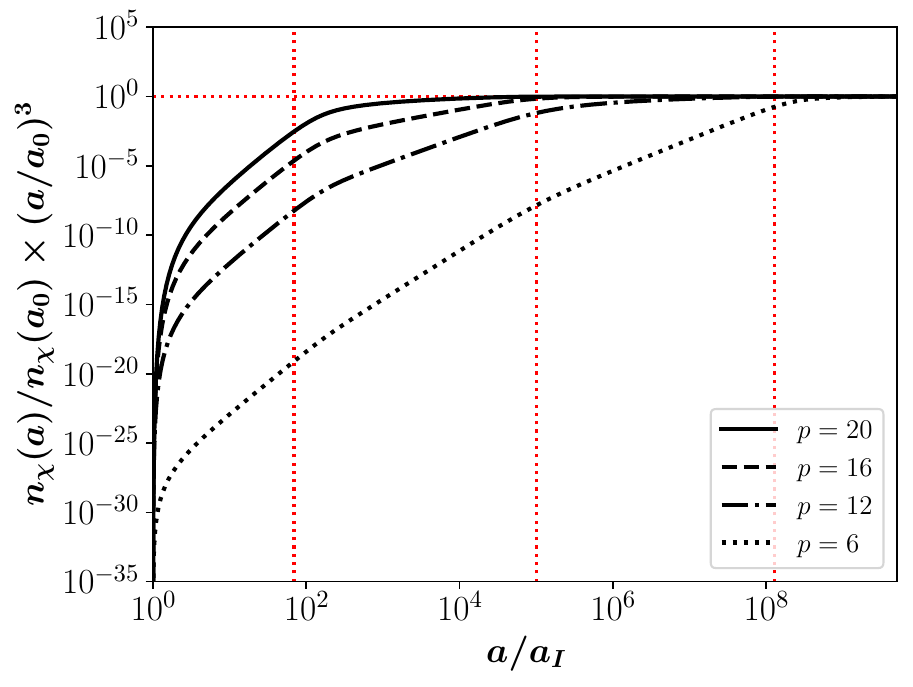}
     \includegraphics[width=\sepf\columnwidth]{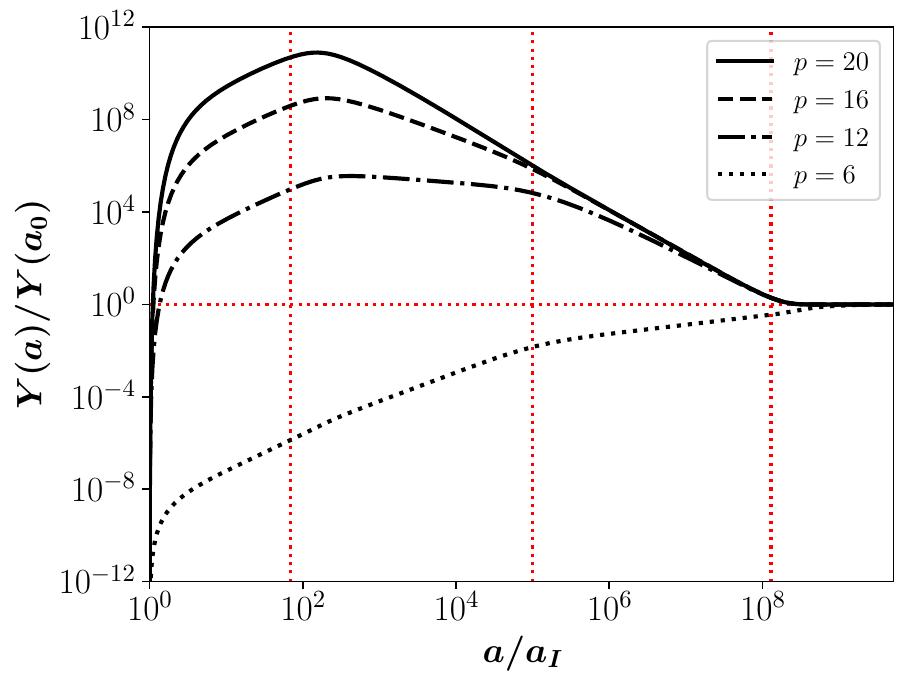}
     \caption{Evolution of the DM comoving number density $n_\chi\, a^3$ (left) and the DM yield $Y$ (right), normalized to their values at late times, for different values of $p$ and the same cosmological parameters used in Fig.~\ref{fig:evoA}.}
     \label{fig:yield}
 \end{figure} 
%%%%%%%%%%%%%%%%%%%%%%%%%%%%%%%%%%%%%%%%%%%%%%%%%%%
Figure~\ref{fig:yield} shows the evolution of the DM comoving number density $n_\chi\, a^3$ (left) and the DM yield $Y$ (right), normalized to their values at later times, for different values of $p$. For the background, the same parameters used in Fig.~\ref{fig:evoA} were used; that is, $H_I = 10^9$~GeV, $m_\phi = 3.3 \times 10^9$~GeV, $m_N = 1.7 \times 10^6$~GeV, $\Gp = 2.5 \times 10^6$~GeV and $\Gamma_N = 3.1\times 10^{-5}$~GeV. While for low values of $p$ DM is mainly produced at the end of reheating, near $T = \Trh$, for increasing values of $p$ the DM production occurs deep during reheating, during the RHN domination. However, DM produced during reheating suffers from a large dilution due to the entropy injected by the decay of the RHNs.

%%%%%%%%%%%%%%%%%%%%%%%%%%%%%%%%%%%%%%%%%%%%%%%%%%%
 \begin{figure}[t!]
     \def\sepf{0.49}
     \centering
     \includegraphics[width=\sepf\columnwidth]{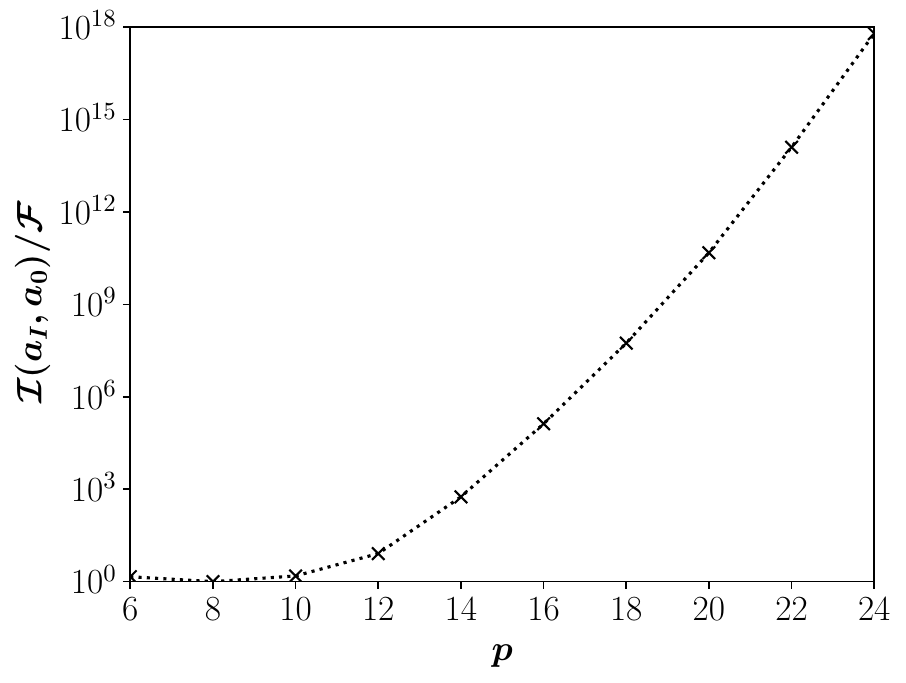}
     \includegraphics[width=\sepf\columnwidth]{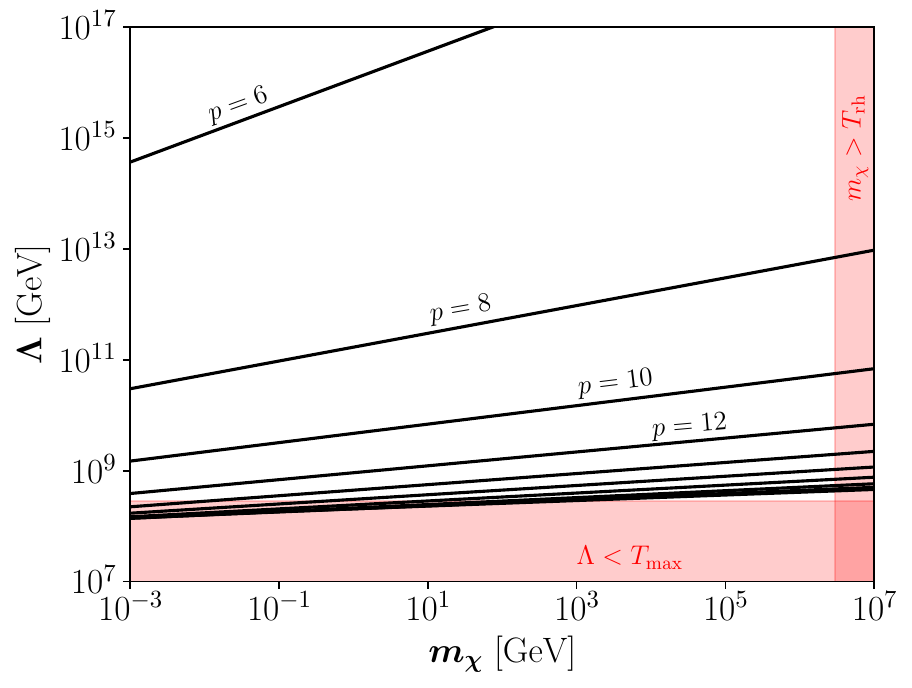}
     \caption{Left: Values of the ratio $\mathcal{I}(a_I,a_0) / \mathcal{F}$ as a function of $p$. Right: scale $\Lambda$ required to fit the whole observed DM abundance as a function of the DM mass. The black lines correspond to even values of $p$ between 6 and 24, while the red bands are outside the validity regime of our calculations because $\Lambda < \Tmax$ or $m_\chi > \Trh$. In both panels, the cosmological parameters were fixed as in Fig.~\ref{fig:evoA}.}
     \label{fig:DM}
 \end{figure} 
%%%%%%%%%%%%%%%%%%%%%%%%%%%%%%%%%%%%%%%%%%%%%%%%%%%
Continuing with the same cosmological parameters fixed in Fig.~\ref{fig:evoA}, the left panel of Fig.~\ref{fig:DM} shows the values of the ratio $\mathcal{I}(a_I,a_0) / \mathcal{F}$ as a function of $p$ (see Table~\ref{tab:I}). The right panel shows instead the scale $\Lambda$ required to reproduce the observed DM abundance as a function of the DM mass. The black lines correspond to even values of $p$ between 6 and 24, while the red bands denoted regions where our calculations are not valid $\Lambda < \Tmax$ (lower band) or $m_\chi > \Trh$ (right band). For a fixed DM mass, smaller values of $p$ require a larger scale $\Lambda$ to reproduce the observed abundance.

%%%%%%%%%%%%%%%%%%%%%%%%%%%%%%%%%%%%%%%%%%%%%%%%%%%
 \begin{figure}[t!]
     \def\sepf{0.49}
     \centering
     \includegraphics[width=\sepf\columnwidth]{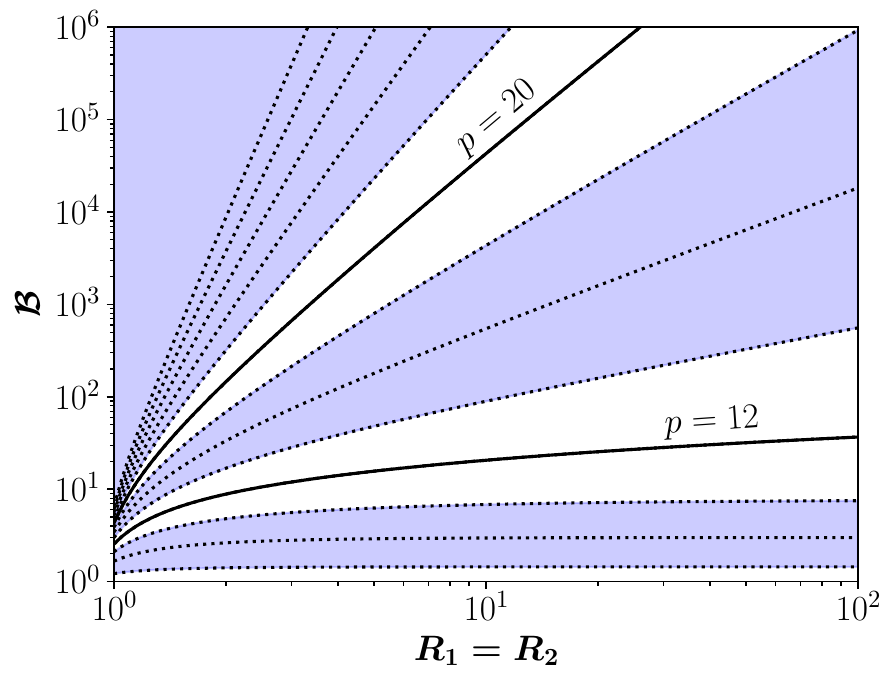}
     \includegraphics[width=\sepf\columnwidth]{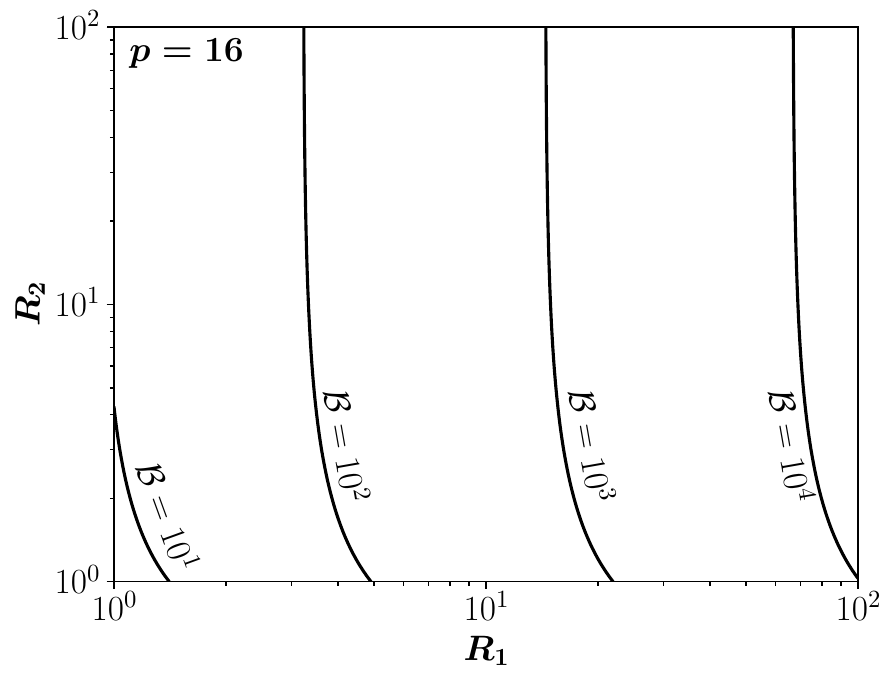}
     \caption{Ratio $\mathcal{B}$ of the total DM produced normalized to what is produced after the end of the reheating. Left panel: the black lines from bottom to top correspond to even values of $p$, increasing from 6 to 30, for $R_1 = R_2$. Right panel: contours of constant $\mathcal{B}$, for $p = 16$.}
     \label{fig:boost}
 \end{figure} 
%%%%%%%%%%%%%%%%%%%%%%%%%%%%%%%%%%%%%%%%%%%%%%%%%%%
To emphasize the importance of seesaw cosmological evolution in DM genesis, Fig.~\ref{fig:boost} shows the ratio $\mathcal{B}$ of the total DM produced normalized to what is produced after the end of the reheating~\cite{Garcia:2017tuj, Bernal:2019mhf, Bernal:2025fdr}
\begin{equation}
    \mathcal{B} = \frac{\mathcal{I}(a_I,a_0)}{\mathcal{I}(\arh,a_0)}\,.
\end{equation}
In the left panel, the black lines from bottom to top correspond to even values of $p$, increasing from 6 to 30, as a function of $R_1 = R_2$. The two solid lines show $p=12$ and $p=20$, while the blue bands depict the ranges $6 \leq p \leq 10$, $14 \leq p \leq 18$, and $22 \leq p$, as described in Table~\ref{tab:I}. For $p = 6$ (which corresponds to an operator of dimension $D = 5$), $\mathcal{B} \simeq 1$ reflects that most DM is produced after the end of reheating, during the SM-dominated era. In contrast, for $p \geq 8$, $\mathcal{B}$ exceeds two once the intermediate RHN-dominated eras are sufficiently long. In addition, in the range $6 \leq p \leq 10$ the ratio $\mathcal{B}$ tends to be independent of $R_1$ and $R_2$, as DM production peaks around $T = \Trh$. From $p =12$, DM is produced mainly deep during reheating and therefore shows a strong dependence on the duration of the RHN dominated era. In the right panel of Fig.~\ref{fig:boost}, contours for constant $\mathcal{B}$ are shown for $p = 16$. Even if $\mathcal{B}$ is a function of $R_1$ and $R_2$, the dependence on $R_2$ is weaker because it reflects the production of DM near $a = \aph$, which is, therefore, very diluted.

%%%%%%%%%%%%%%%%%%%%%%%%%%%%%%%%%%%%%%%%%%%%%%%%%%%%%%%%%%
\section{Conclusions} \label{sec:concl}
%%%%%%%%%%%%%%%%%%%%%%%%%%%%%%%%%%%%%%%%%%%%%%%%%%%%%%%%%%
In this work, we have studied a reheating scenario in which the inflaton transfers its energy to the Standard Model through right-handed neutrinos (RHNs). Since the same particles generate active-neutrino masses through the type-I seesaw mechanism, this setup connects the pre-BBN thermal history to neutrino physics. In particular, neutrino data provide lower bounds on the RHN decay widths and therefore constrain the reheating dynamics.

When the inflaton is much heavier than the RHNs, the latter are produced relativistically and may remain so for a significant period. We have followed their continuously produced nonthermal distribution, accounting for the different momenta and time-dilated decay rates of particles produced at different epochs. For sufficiently separated scales, the Universe passes successively through inflaton domination, relativistic-RHN domination, nonrelativistic-RHN domination, and finally Standard Model radiation domination. The effective equation of state therefore alternates between matter-like and radiation-like behavior.

This evolution leads to a distinctive temperature history. After a rapid initial rise, the Standard Model temperature approaches an approximately constant plateau during inflaton domination. It subsequently scales as $a^{-1/4}$ during relativistic-RHN domination and as $a^{-3/8}$ once the RHNs become nonrelativistic, before recovering the standard adiabatic behavior $T\propto a^{-1}$. The maximum temperature depends on both the inflaton and RHN decay widths, whereas the temperature at the end of reheating is mainly controlled by the RHN lifetime.

We have also investigated the consequences for dark-matter production. For dark matter produced directly in a subdominant inflaton decay channel, the final abundance is determined by the entropy subsequently released through RHN decays. In the hierarchical regime, the yield is controlled by the RHN mass rather than the inflaton mass and can be enhanced relative to conventional reheating by a factor of order $m_\phi/(2m_N)$, at fixed reheating temperature and inflaton branching fraction. For ultraviolet freeze-in, the nonstandard temperature evolution introduces the critical powers $p = 12$ and $20$, which determine the dominant production epoch. The intermediate RHN-dominated eras can consequently produce substantial enhancements of the dark-matter abundance.

Our results demonstrate that the relativistic-to-nonrelativistic transition of an intermediate, long-lived population can qualitatively modify both the expansion and thermal histories of the Universe. The formalism developed here can be extended to multiple nondegenerate RHNs, finite-temperature interaction rates, nonthermal leptogenesis, and other cosmological relics. Seesaw cosmology therefore provides a simple framework linking neutrino-mass generation, reheating, and other pre-BBN cosmic events such as baryogenesis and the origin of dark matter.

%%%%%%%%%%%%%%%%%%%%%%%%%%%%%%%%%%%%%%%%%%%%%%%%%%%%%%%%%%
\section*{Note added}
%%%%%%%%%%%%%%%%%%%%%%%%%%%%%%%%%%%%%%%%%%%%%%%%%%%%%%%%%%
During the finalization of our paper, a similar model was proposed relating the seesaw mechanism with cosmology~\cite{Mambrini:2026tla}, but with an additional scalar mediator, assuming that RHNs thermalize promptly and using a characteristic-momentum approximation rather than the full production-time distribution. Our Eq.~\eqref{eq:Trh} agrees with their main result by replacing the RHN decay width by the scalar mediator width.

%%%%%%%%%%%%%%%%%%%%%%%%%%%%%%%%%%%%%%%%%%%
\acknowledgments
%%%%%%%%%%%%%%%%%%%%%%%%%%%%%%%%%%%%%%%%%%%
The authors acknowledge support from the ICTP through the Associates Programme, where this work was initiated during their visit to ICTP-Trieste in 2023. The authors thank Aoife Bharucha,  Catarina Cosme, and João Rosa for useful discussions. NB received funding from the grants PID2023-151418NB-I00 funded by MCIU/AEI/10.13039/501100011033/ FEDER and PID2022-139841NB-I00 of MICIU/AEI/10.13039/501100011033 and FEDER, UE. OZ has been partially supported by Sostenibilidad-UdeA and the UdeA/ CODI Grants 2022-52380 and  2024-76476. CSF acknowledges the support by Fundação de Amparo à Pesquisa do Estado de São Paulo (FAPESP) Contract No. 2019/11197-6  and Conselho Nacional de Desenvolvimento Científico e Tecnológico (CNPq) under Contract No. 304917/2023-0. CSF and OZ acknowledge support from the ICTP through the Associates Programme (2023-2028).

%%%%%%%%%%%%%%%%%%%%%%%%%%%%%%%%%%%%%%%%%%%%%%%%%%%
\appendix
%%%%%%%%%%%%%%%%%%%%%%%%%%%%%%%%%%%%%%%%%%%%%%%%%%%
%%%%%%%%%%%%%%%%%%%%%%%%%%%%%%%%%%%%%%%%%%%%%%%%%%%
\section{Comparison between the numerical and the analytical approximations} \label{sec:app}
%%%%%%%%%%%%%%%%%%%%%%%%%%%%%%%%%%%%%%%%%%%%%%%%%%%
%%%%%%%%%%%%%%%%%%%%%%%%%%%%%%%%%%%%%%%%%%%%%%%%%%%
 \begin{figure}[b!]
     \def\sepf{0.49}
     \centering
     \includegraphics[width=\sepf\columnwidth]{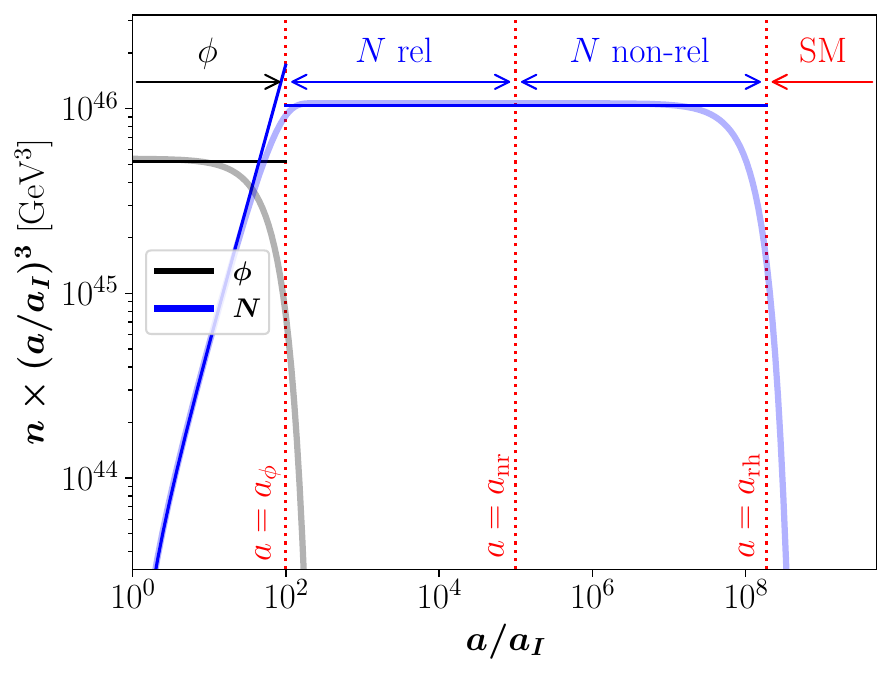}
     \includegraphics[width=\sepf\columnwidth]{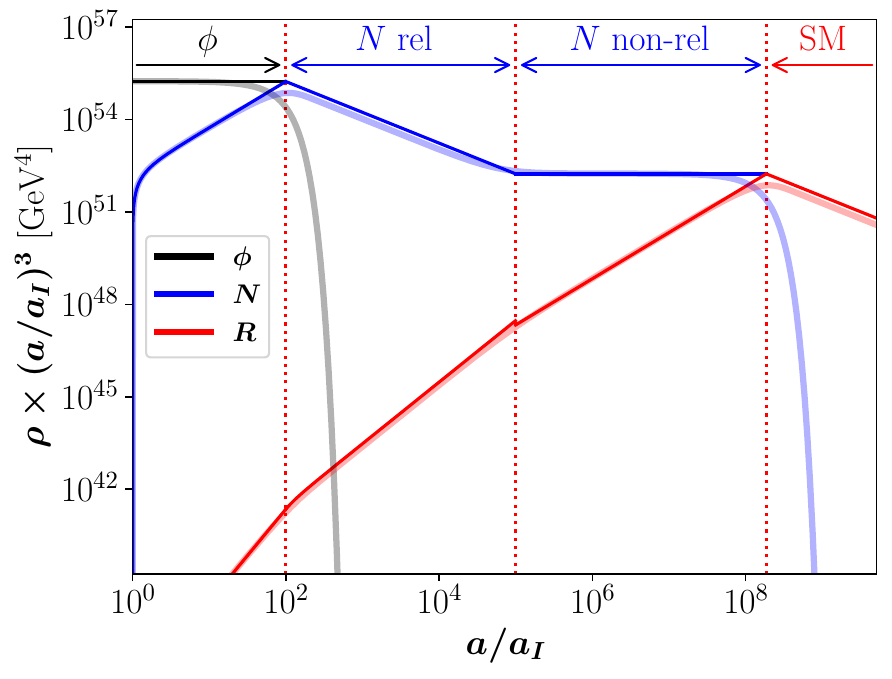}
     \includegraphics[width=\sepf\columnwidth]{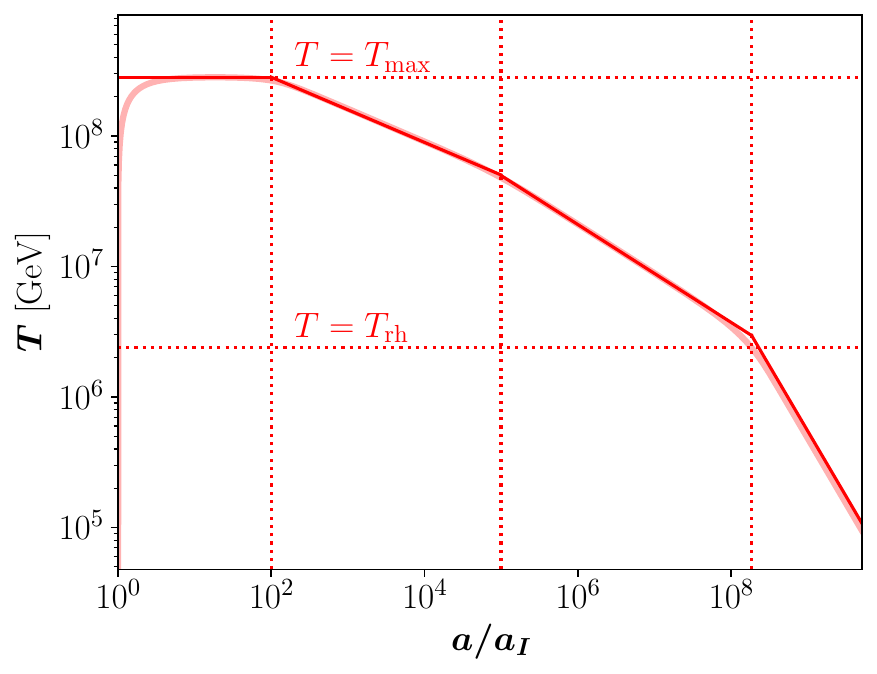}
     \includegraphics[width=\sepf\columnwidth]{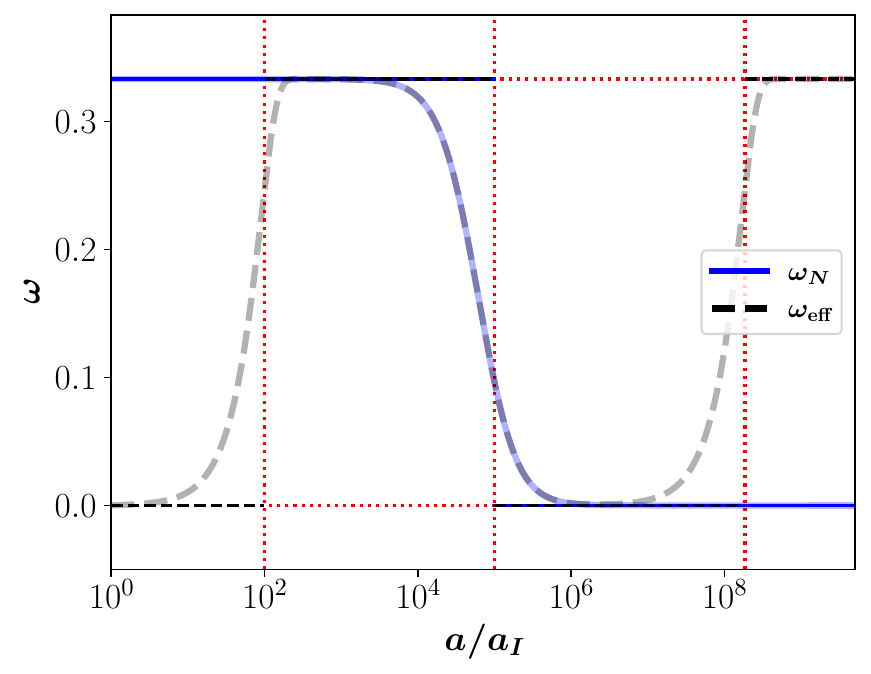}
     \caption{Comparison between the numerical and the analytical approximations, for the benchmark used in Fig.~\ref{fig:evoA}. The thick semitransparent lines correspond to the numerical results, while the thin opaque lines correspond to the analytical expression reported in Section~\ref{sec:anal}.}
     \label{fig:evoA-compa}
 \end{figure} 
%%%%%%%%%%%%%%%%%%%%%%%%%%%%%%%%%%%%%%%%%%%%%%%%%%%
In this appendix, we compare the numerical results with the analytic approximations derived in Section~\ref{sec:anal}. Figure~\ref{fig:evoA-compa} reproduces the numerical results of Fig.~\ref{fig:evoA} as thick semitransparent lines and overlays the corresponding analytic predictions as thin opaque lines. Overall, the agreement is very good, particularly once the system reaches the corresponding asymptotic regimes. As expected, visible deviations arise near the transition scales $a=\aph$, $a=\anr$, and $a=\arh$, where the assumptions underlying the piecewise analytic treatment become less accurate. We emphasize that the analytic expressions rely on the well-separated hierarchy $a_I\ll\aph\ll\anr\ll\arh$. When this hierarchy is not realized, the asymptotic approximations are no longer reliable, and the full numerical solution must be used.

%%%%%%%%%%%%%%%%%%%%%%%%%
\bibliographystyle{JHEP}
\bibliography{biblio}

@article{Mambrini:2026tla,
    author = "Mambrini, Yann",
    title = "{Seesaw reheating}",
    eprint = "2607.08828",
    archivePrefix = "arXiv",
    primaryClass = "hep-ph",
    month = "7",
    year = "2026"
}

@article{Bernal:2021qrl,
    author = "Bernal, Nicol\'as and Xu, Yong",
    title = "{Polynomial inflation and dark matter}",
    eprint = "2106.03950",
    archivePrefix = "arXiv",
    primaryClass = "hep-ph",
    reportNumber = "PI/UAN-2021-690FT",
    doi = "10.1140/epjc/s10052-021-09694-5",
    journal = "Eur. Phys. J. C",
    volume = "81",
    number = "10",
    pages = "877",
    year = "2021"
}

@article{Planck:2018vyg,
    author = "Aghanim, N. and others",
    collaboration = "Planck",
    title = "{Planck 2018 results. VI. Cosmological parameters}",
    eprint = "1807.06209",
    archivePrefix = "arXiv",
    primaryClass = "astro-ph.CO",
    doi = "10.1051/0004-6361/201833910",
    journal = "Astron. Astrophys.",
    volume = "641",
    pages = "A6",
    year = "2020",
    note = "[Erratum: Astron.Astrophys. 652, C4 (2021)]"
}

@article{Bernal:2017kxu,
    author = "Bernal, Nicol\'as and Heikinheimo, Matti and Tenkanen, Tommi and Tuominen, Kimmo and Vaskonen, Ville",
    title = "{The Dawn of FIMP Dark Matter: A Review of Models and Constraints}",
    eprint = "1706.07442",
    archivePrefix = "arXiv",
    primaryClass = "hep-ph",
    reportNumber = "PI-UAN-2017-602FT, HIP-2017-08-TH, PI-UAN--2017--602FT, HIP--2017--08-TH",
    doi = "10.1142/S0217751X1730023X",
    journal = "Int. J. Mod. Phys. A",
    volume = "32",
    number = "27",
    pages = "1730023",
    year = "2017"
}

@article{BICEP:2021xfz,
    author = "Ade, P. A. R. and others",
    collaboration = "BICEP, Keck",
    title = "{Improved Constraints on Primordial Gravitational Waves using Planck, WMAP, and BICEP/Keck Observations through the 2018 Observing Season}",
    eprint = "2110.00483",
    archivePrefix = "arXiv",
    primaryClass = "astro-ph.CO",
    doi = "10.1103/PhysRevLett.127.151301",
    journal = "Phys. Rev. Lett.",
    volume = "127",
    number = "15",
    pages = "151301",
    year = "2021"
}

@article{Ghoshal:2022kqp,
    author = "Ghoshal, Anish and Samanta, Rome and White, Graham",
    title = "{Bremsstrahlung high-frequency gravitational wave signatures of high-scale nonthermal leptogenesis}",
    eprint = "2211.10433",
    archivePrefix = "arXiv",
    primaryClass = "hep-ph",
    doi = "10.1103/PhysRevD.108.035019",
    journal = "Phys. Rev. D",
    volume = "108",
    number = "3",
    pages = "035019",
    year = "2023"
}

@article{Kofman:1997yn,
    author = "Kofman, Lev and Linde, Andrei D. and Starobinsky, Alexei A.",
    title = "{Towards the theory of reheating after inflation}",
    eprint = "hep-ph/9704452",
    archivePrefix = "arXiv",
    reportNumber = "IFA-97-28, SU-ITP-97-18",
    doi = "10.1103/PhysRevD.56.3258",
    journal = "Phys. Rev. D",
    volume = "56",
    pages = "3258--3295",
    year = "1997"
}

@article{Giudice:2000ex,
    author = "Giudice, Gian Francesco and Kolb, Edward W. and Riotto, Antonio",
    title = "{Largest temperature of the radiation era and its cosmological implications}",
    eprint = "hep-ph/0005123",
    archivePrefix = "arXiv",
    reportNumber = "SNS-PH-00-05, FERMILAB-PUB-00-075-A, CERN-TH-2000-107",
    doi = "10.1103/PhysRevD.64.023508",
    journal = "Phys. Rev. D",
    volume = "64",
    pages = "023508",
    year = "2001"
}

@article{Hannestad:2004px,
    author = "Hannestad, Steen",
    title = "{What is the lowest possible reheating temperature?}",
    eprint = "astro-ph/0403291",
    archivePrefix = "arXiv",
    doi = "10.1103/PhysRevD.70.043506",
    journal = "Phys. Rev. D",
    volume = "70",
    pages = "043506",
    year = "2004"
}

@article{Sarkar:1995dd,
    author = "Sarkar, Subir",
    title = "{Big bang nucleosynthesis and physics beyond the standard model}",
    eprint = "hep-ph/9602260",
    archivePrefix = "arXiv",
    reportNumber = "OUTP-95-16-P",
    doi = "10.1088/0034-4885/59/12/001",
    journal = "Rept. Prog. Phys.",
    volume = "59",
    pages = "1493--1610",
    year = "1996"
}

@article{Kawasaki:2000en,
    author = "Kawasaki, M. and Kohri, Kazunori and Sugiyama, Naoshi",
    title = "{MeV scale reheating temperature and thermalization of neutrino background}",
    eprint = "astro-ph/0002127",
    archivePrefix = "arXiv",
    doi = "10.1103/PhysRevD.62.023506",
    journal = "Phys. Rev. D",
    volume = "62",
    pages = "023506",
    year = "2000"
}

@article{Ringwald:2020ist,
    author = {Ringwald, Andreas and Sch\"utte-Engel, Jan and Tamarit, Carlos},
    title = "{Gravitational Waves as a Big Bang Thermometer}",
    eprint = "2011.04731",
    archivePrefix = "arXiv",
    primaryClass = "hep-ph",
    reportNumber = "DESY 20-187, DESY-20-187, TUM-HEP-1293-20",
    doi = "10.1088/1475-7516/2021/03/054",
    journal = "JCAP",
    volume = "03",
    pages = "054",
    year = "2021"
}

@article{Bernal:2019mhf,
    author = "Bernal, Nicol\'as and Elahi, Fatemeh and Maldonado, Carlos and Unwin, James",
    title = "{Ultraviolet Freeze-in and Non-Standard Cosmologies}",
    eprint = "1909.07992",
    archivePrefix = "arXiv",
    primaryClass = "hep-ph",
    reportNumber = "PI/UAN-2019-654FT",
    doi = "10.1088/1475-7516/2019/11/026",
    journal = "JCAP",
    volume = "11",
    pages = "026",
    year = "2019"
}

@article{Barman:2021ugy,
    author = "Barman, Basabendu and Bernal, Nicol\'as",
    title = "{Gravitational SIMPs}",
    eprint = "2104.10699",
    archivePrefix = "arXiv",
    primaryClass = "hep-ph",
    reportNumber = "PI/UAN-2021-688FT",
    doi = "10.1088/1475-7516/2021/06/011",
    journal = "JCAP",
    volume = "06",
    pages = "011",
    year = "2021"
}

@article{Barman:2023ymn,
    author = "Barman, Basabendu and Bernal, Nicol{\'a}s and Xu, Yong and Zapata, {\'O}scar",
    title = "{Gravitational wave from graviton Bremsstrahlung during reheating}",
    eprint = "2301.11345",
    archivePrefix = "arXiv",
    primaryClass = "hep-ph",
    doi = "10.1088/1475-7516/2023/05/019",
    journal = "JCAP",
    volume = "05",
    pages = "019",
    year = "2023"
}

@article{Buchmuller:2011mw,
    author = "Buchmuller, W. and Schmitz, K. and Vertongen, G.",
    title = "{Entropy, Baryon Asymmetry and Dark Matter from Heavy Neutrino Decays}",
    eprint = "1104.2750",
    archivePrefix = "arXiv",
    primaryClass = "hep-ph",
    reportNumber = "DESY-11-002",
    doi = "10.1016/j.nuclphysb.2011.06.004",
    journal = "Nucl. Phys. B",
    volume = "851",
    pages = "481--532",
    year = "2011"
}

@article{Haque:2023zhb,
    author = "Haque, Md Riajul and Maity, Debaprasad and Mondal, Rajesh",
    title = "{Gravitational neutrino reheating}",
    eprint = "2311.07684",
    archivePrefix = "arXiv",
    primaryClass = "hep-ph",
    doi = "10.1103/PhysRevD.109.063543",
    journal = "Phys. Rev. D",
    volume = "109",
    number = "6",
    pages = "063543",
    year = "2024"
}

@article{Cosme:2024ndc,
    author = "Cosme, Catarina and Costa, Francesco and Lebedev, Oleg",
    title = "{Temperature evolution in the Early Universe and freeze-in at stronger coupling}",
    eprint = "2402.04743",
    archivePrefix = "arXiv",
    primaryClass = "hep-ph",
    doi = "10.1088/1475-7516/2024/06/031",
    journal = "JCAP",
    volume = "06",
    pages = "031",
    year = "2024"
}

@article{Hahn-Woernle:2008tsk,
    author = "Hahn-Woernle, F. and Plumacher, M.",
    title = "{Effects of reheating on leptogenesis}",
    eprint = "0801.3972",
    archivePrefix = "arXiv",
    primaryClass = "hep-ph",
    doi = "10.1016/j.nuclphysb.2008.07.032",
    journal = "Nucl. Phys. B",
    volume = "806",
    pages = "68--83",
    year = "2009"
}

@article{Lazarides:1990huy,
    author = "Lazarides, George and Shafi, Q.",
    title = "{Origin of matter in the inflationary cosmology}",
    reportNumber = "BA-90-78",
    doi = "10.1016/0370-2693(91)91090-I",
    journal = "Phys. Lett. B",
    volume = "258",
    pages = "305--309",
    year = "1991"
}

@article{Asaka:1999yd,
    author = "Asaka, T. and Hamaguchi, Koichi and Kawasaki, M. and Yanagida, T.",
    title = "{Leptogenesis in inflaton decay}",
    eprint = "hep-ph/9906366",
    archivePrefix = "arXiv",
    reportNumber = "UT-853, RESCEU-15-99",
    doi = "10.1016/S0370-2693(99)01020-5",
    journal = "Phys. Lett. B",
    volume = "464",
    pages = "12--18",
    year = "1999"
}

@article{Haque:2024zdq,
    author = "Haque, Md Riajul and Maity, Debaprasad and Mondal, Rajesh",
    title = "{Thermal and nonthermal dark matters with gravitational neutrino reheating}",
    eprint = "2408.12450",
    archivePrefix = "arXiv",
    primaryClass = "hep-ph",
    doi = "10.1103/PhysRevD.111.063546",
    journal = "Phys. Rev. D",
    volume = "111",
    number = "6",
    pages = "063546",
    year = "2025"
}

@article{You:2024hit,
    author = "You, Jingtao and Song, Linghao and He, Hong-Jian and Han, Chengcheng",
    title = "{Cosmological non-Gaussianity from the neutrino seesaw mechanism}",
    eprint = "2412.16033",
    archivePrefix = "arXiv",
    primaryClass = "hep-ph",
    doi = "10.1103/lhmk-ycfr",
    journal = "Phys. Rev. D",
    volume = "112",
    number = "8",
    pages = "083555",
    year = "2025"
}

@article{Barman:2023rpg,
    author = "Barman, Basabendu and Bernal, Nicol\'as and Xu, Yong and Zapata, {\'O}scar",
    title = "{Bremsstrahlung-induced gravitational waves in monomial potentials during reheating}",
    eprint = "2305.16388",
    archivePrefix = "arXiv",
    primaryClass = "hep-ph",
    doi = "10.1103/PhysRevD.108.083524",
    journal = "Phys. Rev. D",
    volume = "108",
    number = "8",
    pages = "083524",
    year = "2023"
}

@article{Coy:2024itg,
    author = "Coy, Rupert and Kimus, Jean and Tytgat, Michel H. G.",
    title = "{Light from darkness: history of a hot dark sector}",
    eprint = "2405.10792",
    archivePrefix = "arXiv",
    primaryClass = "hep-ph",
    reportNumber = "ULB-TH/24-02",
    doi = "10.1088/1475-7516/2025/02/077",
    journal = "JCAP",
    volume = "02",
    pages = "077",
    year = "2025"
}

@article{ParticleDataGroup:2024cfk,
    author = "Navas, S. and others",
    collaboration = "Particle Data Group",
    title = "{Review of particle physics}",
    doi = "10.1103/PhysRevD.110.030001",
    journal = "Phys. Rev. D",
    volume = "110",
    number = "3",
    pages = "030001",
    year = "2024"
}

@inproceedings{Baumann:2009ds,
    author = "Baumann, Daniel",
    title = "{Inflation}",
    booktitle = "{Theoretical Advanced Study Institute in Elementary Particle Physics}: {Physics of the Large and the Small}",
    eprint = "0907.5424",
    archivePrefix = "arXiv",
    primaryClass = "hep-th",
    reportNumber = "TASI-2009",
    doi = "10.1142/9789814327183_0010",
    pages = "523--686",
    year = "2011"
}

@article{Barbieri:2025moq,
    author = "Barbieri, Nicola and Brinckmann, Thejs and Gariazzo, Stefano and Lattanzi, Massimiliano and Pastor, Sergio and Pisanti, Ofelia",
    title = "{Current Constraints on Cosmological Scenarios with Very Low Reheating Temperatures}",
    eprint = "2501.01369",
    archivePrefix = "arXiv",
    primaryClass = "astro-ph.CO",
    doi = "10.1103/j5rj-dz1k",
    journal = "Phys. Rev. Lett.",
    volume = "135",
    number = "18",
    pages = "181003",
    year = "2025"
}

@article{Traschen:1990sw,
    author = "Traschen, Jennie H. and Brandenberger, Robert H.",
    title = "{Particle Production During Out-of-equilibrium Phase Transitions}",
    reportNumber = "BROWN-HET-731",
    doi = "10.1103/PhysRevD.42.2491",
    journal = "Phys. Rev. D",
    volume = "42",
    pages = "2491--2504",
    year = "1990"
}

@article{Kofman:1994rk,
    author = "Kofman, Lev and Linde, Andrei D. and Starobinsky, Alexei A.",
    title = "{Reheating after inflation}",
    eprint = "hep-th/9405187",
    archivePrefix = "arXiv",
    reportNumber = "UH-IFA-94-35, SU-ITP-94-13, YITP-U-94-15",
    doi = "10.1103/PhysRevLett.73.3195",
    journal = "Phys. Rev. Lett.",
    volume = "73",
    pages = "3195--3198",
    year = "1994"
}

@article{Dolgov:1989us,
    author = "Dolgov, A. D. and Kirilova, D. P.",
    title = "{On Particle Creation by a Time Dependent Scalar Field}",
    reportNumber = "JINR-E2-89-321",
    journal = "Sov. J. Nucl. Phys.",
    volume = "51",
    pages = "172--177",
    year = "1990"
}

@article{Barman:2025lvk,
    author = "Barman, Basabendu and Bernal, Nicol{\'a}s and Rubio, Javier",
    title = "{Two or three things particle physicists (mis)understand about (pre)heating}",
    eprint = "2503.19980",
    archivePrefix = "arXiv",
    primaryClass = "hep-ph",
    reportNumber = "IPARCOS-UCM-25-019",
    doi = "10.1016/j.nuclphysb.2025.116996",
    journal = "Nucl. Phys. B",
    volume = "1018",
    pages = "116996",
    year = "2025"
}

@article{Datta:2025wfh,
    author = "Datta, Arghyajit and Khalil, Shaaban and Mandal, Rajat Kumar and Sil, Arunansu",
    title = "{Probing right handed neutrino assisted reheating with gravitational waves and leptogenesis}",
    eprint = "2507.09728",
    archivePrefix = "arXiv",
    primaryClass = "hep-ph",
    doi = "10.1088/1475-7516/2026/02/061",
    journal = "JCAP",
    volume = "02",
    pages = "061",
    year = "2026"
}

@article{Borboruah:2025hai,
    author = "Borboruah, Zafri A. and Deppisch, Frank F. and Ghoshal, Anish and Malhotra, Lekhika",
    title = "{Inflationary gravitational waves and laboratory searches as complementary probes of right-handed neutrinos}",
    eprint = "2504.15374",
    archivePrefix = "arXiv",
    primaryClass = "hep-ph",
    doi = "10.1103/qyld-mf33",
    journal = "Phys. Rev. D",
    volume = "112",
    number = "5",
    pages = "056003",
    year = "2025"
}

@article{Allahverdi:2020bys,
    author = "Allahverdi, Rouzbeh and others",
    title = "{The First Three Seconds: a Review of Possible Expansion Histories of the Early Universe}",
    eprint = "2006.16182",
    archivePrefix = "arXiv",
    primaryClass = "astro-ph.CO",
    reportNumber = "FERMILAB-PUB-20-242-A, KCL-PH-TH/2020-33, KEK-Cosmo-257,
  KEK-TH-2231, IPMU20-0070, PI/UAN-2020-674FT, RUP-20-22",
    doi = "10.21105/astro.2006.16182",
    journal = "Open J. Astrophys.",
    volume = "4",
    pages = "astro.2006.16182",
    year = "2021"
}

@article{McDonald:2001vt,
    author = "McDonald, John",
    title = "{Thermally generated gauge singlet scalars as selfinteracting dark matter}",
    eprint = "hep-ph/0106249",
    archivePrefix = "arXiv",
    doi = "10.1103/PhysRevLett.88.091304",
    journal = "Phys. Rev. Lett.",
    volume = "88",
    pages = "091304",
    year = "2002"
}

@article{Choi:2005vq,
    author = "Choi, Ki-Young and Roszkowski, Leszek",
    editor = "Choi, Kiwoon and Kim, Jihn E. and Son, Dongchul",
    title = "{E-WIMPs}",
    eprint = "hep-ph/0511003",
    archivePrefix = "arXiv",
    doi = "10.1063/1.2149672",
    journal = "AIP Conf. Proc.",
    volume = "805",
    number = "1",
    pages = "30--36",
    year = "2005"
}

@article{Asaka:1999jb,
    author = "Asaka, T. and Hamaguchi, Koichi and Kawasaki, M. and Yanagida, T.",
    title = "{Leptogenesis in inflationary universe}",
    eprint = "hep-ph/9907559",
    archivePrefix = "arXiv",
    reportNumber = "UT-855, RESCEU-28-99",
    doi = "10.1103/PhysRevD.61.083512",
    journal = "Phys. Rev. D",
    volume = "61",
    pages = "083512",
    year = "2000"
}

@article{Kusenko:2006rh,
    author = "Kusenko, Alexander",
    title = "{Sterile neutrinos, dark matter, and the pulsar velocities in models with a Higgs singlet}",
    eprint = "hep-ph/0609081",
    archivePrefix = "arXiv",
    reportNumber = "UCLA-06-TEP-23",
    doi = "10.1103/PhysRevLett.97.241301",
    journal = "Phys. Rev. Lett.",
    volume = "97",
    pages = "241301",
    year = "2006"
}

@article{Hall:2009bx,
    author = "Hall, Lawrence J. and Jedamzik, Karsten and March-Russell, John and West, Stephen M.",
    title = "{Freeze-In Production of FIMP Dark Matter}",
    eprint = "0911.1120",
    archivePrefix = "arXiv",
    primaryClass = "hep-ph",
    reportNumber = "OUTP-09-18-P, UCB-PTH-09-32",
    doi = "10.1007/JHEP03(2010)080",
    journal = "JHEP",
    volume = "03",
    pages = "080",
    year = "2010"
}

@article{Elahi:2014fsa,
    author = "Elahi, Fatemeh and Kolda, Christopher and Unwin, James",
    title = "{UltraViolet Freeze-in}",
    eprint = "1410.6157",
    archivePrefix = "arXiv",
    primaryClass = "hep-ph",
    doi = "10.1007/JHEP03(2015)048",
    journal = "JHEP",
    volume = "03",
    pages = "048",
    year = "2015"
}

@article{Amin:2014eta,
    author = "Amin, Mustafa A. and Hertzberg, Mark P. and Kaiser, David I. and Karouby, Johanna",
    title = "{Nonperturbative Dynamics Of Reheating After Inflation: A Review}",
    eprint = "1410.3808",
    archivePrefix = "arXiv",
    primaryClass = "hep-ph",
    doi = "10.1142/S0218271815300037",
    journal = "Int. J. Mod. Phys. D",
    volume = "24",
    pages = "1530003",
    year = "2014"
}

@article{Zhang:2023oyo,
    author = "Zhang, Xinyi",
    title = "{Towards a systematic study of non-thermal leptogenesis from inflaton decays}",
    eprint = "2311.05824",
    archivePrefix = "arXiv",
    primaryClass = "hep-ph",
    doi = "10.1007/JHEP05(2024)147",
    journal = "JHEP",
    volume = "05",
    pages = "147",
    year = "2024"
}

@article{Zhang:2025jfh,
    author = "Zhang, Xinyi",
    title = "{Neutrino reheating predictions with non-thermal leptogenesis}",
    eprint = "2501.16114",
    archivePrefix = "arXiv",
    primaryClass = "hep-ph",
    doi = "10.1007/JHEP07(2025)099",
    journal = "JHEP",
    volume = "07",
    pages = "099",
    year = "2025"
}

@article{Petraki:2007gq,
    author = "Petraki, Kalliopi and Kusenko, Alexander",
    title = "{Dark-matter sterile neutrinos in models with a gauge singlet in the Higgs sector}",
    eprint = "0711.4646",
    archivePrefix = "arXiv",
    primaryClass = "hep-ph",
    reportNumber = "UCLA-07-TEP-27",
    doi = "10.1103/PhysRevD.77.065014",
    journal = "Phys. Rev. D",
    volume = "77",
    pages = "065014",
    year = "2008"
}

@article{Garcia:2017tuj,
    author = "García, Marcos A. G. and Mambrini, Yann and Olive, Keith A. and Peloso, Marco",
    title = "{Enhancement of the Dark Matter Abundance Before Reheating: Applications to Gravitino Dark Matter}",
    eprint = "1709.01549",
    archivePrefix = "arXiv",
    primaryClass = "hep-ph",
    reportNumber = "LPT--ORSAY-17-36, UMN--TH--3635-17, FTPI--MINN--17-15",
    doi = "10.1103/PhysRevD.96.103510",
    journal = "Phys. Rev. D",
    volume = "96",
    number = "10",
    pages = "103510",
    year = "2017"
}

@article{Bernal:2025fdr,
    author = "Bernal, Nicol{\'a}s and Deka, Kuldeep and Losada, Marta",
    title = "{Dark matter ultraviolet freeze-in in general reheating scenarios}",
    eprint = "2501.04774",
    archivePrefix = "arXiv",
    primaryClass = "hep-ph",
    doi = "10.1103/PhysRevD.111.055034",
    journal = "Phys. Rev. D",
    volume = "111",
    number = "5",
    pages = "055034",
    year = "2025"
}

@article{Han:2024qbw,
    author = "Han, Chengcheng and He, Hong-Jian and Song, Linghao and You, Jingtao",
    title = "{Cosmological signatures of neutrino seesaw mechanism}",
    eprint = "2412.21045",
    archivePrefix = "arXiv",
    primaryClass = "hep-ph",
    doi = "10.1103/b7hv-3h2p",
    journal = "Phys. Rev. D",
    volume = "112",
    number = "8",
    pages = "L081309",
    year = "2025"
}

@article{Davidson:2008bu,
    author = "Davidson, Sacha and Nardi, Enrico and Nir, Yosef",
    title = "{Leptogenesis}",
    eprint = "0802.2962",
    archivePrefix = "arXiv",
    primaryClass = "hep-ph",
    doi = "10.1016/j.physrep.2008.06.002",
    journal = "Phys. Rept.",
    volume = "466",
    pages = "105--177",
    year = "2008"
}

@article{Fong:2012buy,
    author = "Fong, Chee Sheng and Nardi, Enrico and Riotto, Antonio",
    title = "{Leptogenesis in the Universe}",
    eprint = "1301.3062",
    archivePrefix = "arXiv",
    primaryClass = "hep-ph",
    doi = "10.1155/2012/158303",
    journal = "Adv. High Energy Phys.",
    volume = "2012",
    pages = "158303",
    year = "2012"
}

@article{Jaeckel:2020oet,
    author = "Jaeckel, Joerg and Yin, Wen",
    title = "{Boosted Neutrinos and Relativistic Dark Particles as Messengers from Reheating}",
    eprint = "2007.15006",
    archivePrefix = "arXiv",
    primaryClass = "hep-ph",
    doi = "10.1088/1475-7516/2021/02/044",
    journal = "JCAP",
    volume = "02",
    pages = "044",
    year = "2021"
}

@article{Batell:2024dsi,
    author = "Batell, Brian and others",
    title = "{Conversations and deliberations: Non-standard cosmological epochs and expansion histories}",
    eprint = "2411.04780",
    archivePrefix = "arXiv",
    primaryClass = "astro-ph.CO",
    doi = "10.1142/S0217751X25300042",
    journal = "Int. J. Mod. Phys. A",
    volume = "40",
    number = "17",
    pages = "2530004",
    year = "2025"
}

@article{Casas:2001sr,
    author = "Casas, J. A. and Ibarra, A.",
    title = "{Oscillating neutrinos and $\mu \to e, \gamma$}",
    eprint = "hep-ph/0103065",
    archivePrefix = "arXiv",
    reportNumber = "IEM-FT-211-01, OUTP-01-11P, IFT-UAM-CSIC-01-08",
    doi = "10.1016/S0550-3213(01)00475-8",
    journal = "Nucl. Phys. B",
    volume = "618",
    pages = "171--204",
    year = "2001"
}

@article{Minkowski:1977sc,
    author = "Minkowski, Peter",
    title = "{$\mu \to e\gamma$ at a Rate of One Out of $10^{9}$ Muon Decays?}",
    reportNumber = "Print-77-0182 (BERN)",
    doi = "10.1016/0370-2693(77)90435-X",
    journal = "Phys. Lett. B",
    volume = "67",
    pages = "421--428",
    year = "1977"
}

@article{Yanagida:1979as,
    author = "Yanagida, Tsutomu",
    editor = "Sawada, Osamu and Sugamoto, Akio",
    title = "{Horizontal gauge symmetry and masses of neutrinos}",
    reportNumber = "KEK-79-18-95",
    journal = "Conf. Proc. C",
    volume = "7902131",
    pages = "95--99",
    year = "1979"
}

@article{Glashow:1979nm,
    author = "Glashow, S. L.",
    editor = "L{\'e}vy, Maurice and Basdevant, Jean-Louis and Speiser, David and Weyers, Jacques and Gastmans, Raymond and Jacob, Maurice",
    title = "{The Future of Elementary Particle Physics}",
    reportNumber = "HUTP-79-A059",
    doi = "10.1007/978-1-4684-7197-7_15",
    journal = "NATO Sci. Ser. B",
    volume = "61",
    pages = "687",
    year = "1980"
}

@article{Gell-Mann:1979vob,
    author = "Gell-Mann, Murray and Ramond, Pierre and Slansky, Richard",
    title = "{Complex Spinors and Unified Theories}",
    eprint = "1306.4669",
    archivePrefix = "arXiv",
    primaryClass = "hep-th",
    reportNumber = "PRINT-80-0576",
    journal = "Conf. Proc. C",
    volume = "790927",
    pages = "315--321",
    year = "1979"
}

@article{Mohapatra:1979ia,
    author = "Mohapatra, Rabindra N. and Senjanović, Goran",
    title = "{Neutrino Mass and Spontaneous Parity Nonconservation}",
    reportNumber = "MDDP-TR-80-060, MDDP-PP-80-105, CCNY-HEP-79-10",
    doi = "10.1103/PhysRevLett.44.912",
    journal = "Phys. Rev. Lett.",
    volume = "44",
    pages = "912",
    year = "1980"
}

@article{Bernal:2025lxp,
    author = "Bernal, Nicol{\'a}s and Wu, Quan-feng and Xu, Xun-Jie and Xu, Yong",
    title = "{Pre-thermalized gravitational waves}",
    eprint = "2503.10756",
    archivePrefix = "arXiv",
    primaryClass = "hep-ph",
    reportNumber = "MITP-25-022",
    doi = "10.1007/JHEP08(2025)125",
    journal = "JHEP",
    volume = "08",
    pages = "125",
    year = "2025"
}

@article{Bernal:2023wus,
    author = "Bernal, Nicol{\'a}s and Cl{\'e}ry, Simon and Mambrini, Yann and Xu, Yong",
    title = "{Probing reheating with graviton bremsstrahlung}",
    eprint = "2311.12694",
    archivePrefix = "arXiv",
    primaryClass = "hep-ph",
    reportNumber = "MITP-23-065",
    doi = "10.1088/1475-7516/2024/01/065",
    journal = "JCAP",
    volume = "01",
    pages = "065",
    year = "2024"
}

@inproceedings{Barman:2024htg,
    author = "Barman, Basabendu and Bernal, Nicol{\'a}s and Cl{\'e}ry, Simon and Mambrini, Yann and Xu, Yong and Zapata, {\'O}scar",
    title = "{Probing Reheating with Gravitational Waves from Graviton Bremsstrahlung}",
    booktitle = "{58$^{th}$ Rencontres de Moriond on Electroweak Interactions and Unified Theories}",
    eprint = "2405.09620",
    archivePrefix = "arXiv",
    primaryClass = "astro-ph.CO",
    month = "5",
    year = "2024"
}

@article{Garny:2017kha,
    author = "Garny, Mathias and Palessandro, Andrea and Sandora, McCullen and Sloth, Martin S.",
    title = "{Theory and Phenomenology of Planckian Interacting Massive Particles as Dark Matter}",
    eprint = "1709.09688",
    archivePrefix = "arXiv",
    primaryClass = "hep-ph",
    doi = "10.1088/1475-7516/2018/02/027",
    journal = "JCAP",
    volume = "02",
    pages = "027",
    year = "2018"
}

@article{Bernal:2018qlk,
    author = "Bernal, Nicol{\'a}s and Dutra, Ma{\'\i}ra and Mambrini, Yann and Olive, Keith and Peloso, Marco and Pierre, Mathias",
    title = "{Spin-2 Portal Dark Matter}",
    eprint = "1803.01866",
    archivePrefix = "arXiv",
    primaryClass = "hep-ph",
    doi = "10.1103/PhysRevD.97.115020",
    journal = "Phys. Rev. D",
    volume = "97",
    number = "11",
    pages = "115020",
    year = "2018"
}

@article{Garny:2015sjg,
    author = "Garny, Mathias and Sandora, McCullen and Sloth, Martin S.",
    title = "{Planckian Interacting Massive Particles as Dark Matter}",
    eprint = "1511.03278",
    archivePrefix = "arXiv",
    primaryClass = "hep-ph",
    reportNumber = "CERN-PH-TH-2015-264",
    doi = "10.1103/PhysRevLett.116.101302",
    journal = "Phys. Rev. Lett.",
    volume = "116",
    number = "10",
    pages = "101302",
    year = "2016"
}

@article{Tang:2017hvq,
    author = "Tang, Yong and Wu, Yue-Liang",
    title = "{On Thermal Gravitational Contribution to Particle Production and Dark Matter}",
    eprint = "1708.05138",
    archivePrefix = "arXiv",
    primaryClass = "hep-ph",
    reportNumber = "UT-17-27",
    doi = "10.1016/j.physletb.2017.10.034",
    journal = "Phys. Lett. B",
    volume = "774",
    pages = "676--681",
    year = "2017"
}

@article{Cline:2026jra,
    author = "Cline, James M. and Xu, Yong",
    title = "{Irreducible Graviton Floor from Reheating}",
    eprint = "2605.16201",
    archivePrefix = "arXiv",
    primaryClass = "hep-ph",
    month = "5",
    year = "2026"
}

@article{Esteban:2024eli,
    author = "Esteban, Ivan and González-García, M. C. and Maltoni, Michele and Martínez-Soler, Ivan and Pinheiro, Jo{\~a}o Paulo and Schwetz, Thomas",
    title = "{NuFit-6.0: updated global analysis of three-flavor neutrino oscillations}",
    eprint = "2410.05380",
    archivePrefix = "arXiv",
    primaryClass = "hep-ph",
    reportNumber = "IFT-UAM/CSIC-24-140, YITP-SB-2024-24, IPPP/24/64, IPPP/24/64, IFT-UAM/CSIC-24-140, YITP-SB-2024-24",
    doi = "10.1007/JHEP12(2024)216",
    journal = "JHEP",
    volume = "12",
    pages = "216",
    year = "2024"
}

@article{deSalas:2020pgw,
    author = "de Salas, P. F. and Forero, D. V. and Gariazzo, S. and Mart{\'\i}nez-Mirav{\'e}, P. and Mena, O. and Ternes, C. A. and T{\'o}rtola, M. and Valle, J. W. F.",
    title = "{2020 global reassessment of the neutrino oscillation picture}",
    eprint = "2006.11237",
    archivePrefix = "arXiv",
    primaryClass = "hep-ph",
    doi = "10.1007/JHEP02(2021)071",
    journal = "JHEP",
    volume = "02",
    pages = "071",
    year = "2021"
}

@article{Fukugita:1986hr,
    author = "Fukugita, M. and Yanagida, T.",
    title = "{Baryogenesis Without Grand Unification}",
    reportNumber = "RIFP-641",
    doi = "10.1016/0370-2693(86)91126-3",
    journal = "Phys. Lett. B",
    volume = "174",
    pages = "45--47",
    year = "1986"
}

@article{Bernal:2017zvx,
    author = "Bernal, Nicol{\'a}s and Fong, Chee Sheng",
    title = "{Hot Leptogenesis from Thermal Dark Matter}",
    eprint = "1707.02988",
    archivePrefix = "arXiv",
    primaryClass = "hep-ph",
    reportNumber = "PI-UAN-2017-608FT, FERMILAB-PUB-17-342-T",
    doi = "10.1088/1475-7516/2017/10/042",
    journal = "JCAP",
    volume = "10",
    pages = "042",
    year = "2017"
}

@article{Bernal:2026ljj,
  author = {Bernal, Nicol{\'a}s and Fong, Chee Sheng},
  title = {{Quantum Seesaw Cosmology: A Heating Phase from Pauli Blocking}},
  eprint = {2609.17681},
  archiveprefix = {arXiv},
  primaryclass = {hep-ph},
  month = {9},
  year = {2026},
}
%%%%%%%%%%%%%%%%%%%%%%%%%
\end{document}